\documentclass[5pt]{aastex61}
\usepackage[utf8]{inputenc}
\usepackage{nameref}
\usepackage{wasysym}
\usepackage{makecell}
\usepackage[caption=false]{subfig}

\usepackage{graphicx}
\usepackage{sidecap}
\usepackage{soul}

\graphicspath{{img/}}

\newcommand{\sitem}{\vspace{-0.3cm}\item}
\begin{document}
	
\title{The distribution of heavy-elements in giant protoplanetary atmospheres: the importance of planetesimal-envelope interactions}
\author{Claudio Valletta}
\affiliation{Center for Theoretical Astrophysics and Cosmology \\
			Institute for Computational Science, University of Zürich \\
			Winterthurerstrasse 190, 8057 Zürich, Switzerland}
\email{valletta@physik.uzh.ch}

\author{Ravit Helled}
\affiliation{Center for Theoretical Astrophysics and Cosmology \\
			Institute for Computational Science, University of Zürich \\
			Winterthurerstrasse 190, 8057 Zürich, Switzerland}

%\correspondingauthor{Simon Müller}

%% Reintroduced the \received and \accepted commands from AASTeX v5.2
\received{September 3, 2018}
\revised{October 16, 2018}
\accepted{November 24, 2018}
%% Command to document which AAS Journal the manuscript was submitted to.
%% Adds "Submitted to " the arguement.
\submitjournal{ApJ}

\begin{abstract}

In the standard model for giant planet formation, the planetary growth begins with accretion of solids followed by a buildup of a gaseous atmosphere as more solids are accreted, and finally, by rapid accretion of gas. The interaction of the solids with the gaseous envelope determines the subsequent planetary growth and the final internal structure. 
%\textbf{Most simulations performed so far always assume that solids reach the core intact, i.e. the envelope has a solar composition.} \textbf{It is known that planetesimal are affected by ablation and fragmentation while entering planet's envelope. Thus, not all planetesimals reach the core intact and the planet’s envelope gets enriched in high-z material.}
%As the heavy elements are accreted, gas can also be accreted and the subsequent growth and the 
%The physical properties of the solids, such as their sizes, composition and material strength, and their interaction with the proto-planetary envelope significantly affect the distribution of the heavy elements within the proto-planet and its consequential growth. 
%\st{Understudying the predicted heavy-element distribution in giant planet envelope is critical for the characterization of Jupiter and gaseous giant exoplanets, and for connecting planet formation theories with internal structure models.} 
In this work we simulate the interaction of planetesimals with a growing giant planet (proto-Jupiter) and investigate how different treatments of the planetesimal-envelope interaction affect the heavy-element distribution, and the inferred core mass. We consider various planetesimal sizes and compositions as well as different ablation and radiation efficiencies, and fragmentation models. We find that in most cases %independently of the planetesimal size and composition 
the core reaches a maximum mass of $\sim$ 2 M$_{\oplus}$. We show that the value of the core's mass mainly depends on the assumed size and composition of the solids, 
while the heavy-element distribution is also affected by the fate of the accreted planetesimals (ablation/fragmentation). 
Fragmentation, which is found to be important for planetesimals $>$ 1 km,  typically leads to enrichment of the inner part of the envelope while ablation results in enrichment of the outer atmosphere. 
Finally, we present a semi-analytical prescription for deriving the heavy-element distribution in giant protoplanets. 
% planetesimal tends to deposit their mass also in the outer part of the atmosphere. We find that planetesimal fragmentation is important for planetesimal larger that 1 km.}Finally, we present a semi-analytical prescription for deriving the heavy-element distribution in giant protoplanets. 
%This can b e easily implemented in giant planet formation calculation and can 
%be used to include the presence of the heavy-elements in a more consistent manner, i.e., accounting for their effect on the EOS and opacity calculation
%%This prediction is strong and is not sensitive to the size or composition of the planetesimal and its interaction with the envelope. 
%We demonstrate the importance of accounting for planetesimal ablation, %as it can change the envelope mass (and density) during the early by a great amount. 
%% in the early stages, the mass of the atmosphere can be 3 - 4 order of magnitude greater than for a pure H-He envelope. This is expected to have dramatic consequence on the subsequent evolution of the planet. 
%%Finally we 
%and present a simple semi-analytical approach to derive the heavy-element distribution in giant protoplanets. 
%

\end{abstract}


\section{Introduction} \label{sec:introduction}
The ongoing  characterization and discoveries of giant exoplanets and the accurate measurements of the giant planets in the Solar System provide a unique opportunity to understand these astronomical objects. 
As more information is available, theoretical models are challenged to explain the observed properties. This is not always an easy task, in particular when we aim to connect giant planetary interiors with giant planet formation (e.g., \citealt{Helled14}).
\par

%Recently, a great effort from the theoretical side has been demonstrated and both formation and structure models of giant planets have been improved. 
%Different aspects have been studied, including the {\bf bla bra - Claudio - read the relevant papers and cite them here!}\\
%By now, it is clear that the simple assumption that the accreted heavy elements reach the core has been abondand and structure Jupiter models allow for a more complex structure, which also affect their predicted evolution history \citep{Venturini16,Lozovsky17}. 
%
%
%final composition of the planet and the initial
%distribution of heavy elements.  
%\st{At the beginning of their formation, giant planets are capable of binding only a very tenuous envelope so that infalling  planetesimals essentially reach the core directly} (see e.g. \cite{Pollack96, Iaroslav07, Ravit06}).
%\textit{I would move these two lines down}
%Understanding the formation mechanism of giant planets is critical for the  for revealing information on the chemical and physical properties of protoplanetary disks where planets form, as well as for the delivery of volatiles to the inner part of the planetary system and the overall dynamical evolution of the system in early stages. 
%Since decades, theoretical models have been developed to simulate the formation of giant planets. 
In the standard model for giant planet formation, core accretion, the planetary growth begins with the formation of a core (\citealt{Alibert05}, \citealt{P96}, \citealt{HelledPPVI} and references therein). 
Once the core reaches about Mars' mass, its gravity is strong enough to bind hydrogen and helium (hereafter H-He) gas from the protoplanetary disk. 
Then, the protoplanet keeps growing by accreting both solids (heavy elements) and H-He until crossover mass is reached and rapid gas accretion takes place. 
In the early core accretion simulations for the sake of numerical simplicity, it was assumed that all the heavy elements reach the core while the envelope is composed of H-He \citep{P96}. During the initial stages of planetary formation, the proto-planet is capable of binding only a very tenuous envelope, so that infalling  planetesimals reach the core directly  (e.g., \citealt{P96}). Once the core mass reaches a small value of $\sim$ 1-2 M$_{\oplus}$ \citep{P86,Brouwers2017,Iaroslav07,Venturini16,Lozovsky17} and is surrounded by gas, solids (planetesimals/pebbles) composed of heavy-elements are expected to dissolve in the envelope instead of reaching the core. 
%As already discussed in various papers \citep{Venturini15,Venturini16,Lozovsky17} the original assumption of P96 of the planetesimals reaching the core was made for a numerical simplification, and already at that time it was observed that once the core mass reaches $\sim$ 1-2 M$_{\oplus}$ planetesimals are \textbf{fully} ablated in the envelope. 
\par

While it is known that the heavy elements can remain in the envelope, their actual distribution is not well constrained. 
Nevertheless, determining the fate of the accreted heavy elements and their distribution within the envelope is important for several reasons. 
First, a non-homogenous internal structure has a significant impact on the thermal evolution and final structure of the planets (\citealt{Lozovsky17},\citealt{Vazan2018}). 
Second, the presence of heavy elements material in the envelope can dramatically affect the consequent growth of the planet (\citealt{HI11},\citealt{Venturini16}).
Finally, the deposition of heavy elements can change the local conditions at the envelope such as the opacity and heat transport mechanism. %in the envelope. 
% and the heat transport in mechanism and significantly affect the stability against convection and mixing.}
%\textit{The following paragraph was in the end of the introduction, I moved here}
%{\bf to be fixed...}\\
Despite its importance, envelope enrichment is often neglected, or being treated in a simple manner.  This is mainly due to the difficulty in following the planetesimal-envelope interaction in detail and at the same time model the subsequent 
planetary growth while accounting for the change in the equation of state and opacity of the envelope due to heavy-element deposition.
These two aspects involve different physical processes, and therefore studies typically concentrate on the interaction between heavy elements and the planetary envelope (e.g., \citealt{Venturini16},\citealt{Lozovsky17}, \citealt{Brouwers2017}) or on the effect of the heavy-elements on planetary growth and long-term evolution (e.g., \citealt{Vazan2018}).

%\st{Given a certain atmospheric model must be calculated the high-Z mass deposition distribution in the atmosphere building a model that describe the incoming planetesimal (its velocity, size, composition, and impact parameter) and modeling its thermal ablation.} In the case of planetesimals another important process that must be described with ablation is the mechanical break-up/fragmentation (\citealt{Podolak88}, \citealt{Mordasini06}).
%In order to predict the distribution of heavy elements in protoplanetary atmospheres the interactions of solids with the protoplanet has to be considered and 
% then it becomes crucial to treat the planetesimal correctly, describing, accurately, the equation of motion, the mass ablation and finally the fragmentation model. Finally, 
%the subsequent planetary growth, accounting for the heavies,  must be included. 
\par

%{\bf more details on the methods and results...}\\
Previous research on planetesimal ablation in giant protoplanets has been mostly focused on the inferred core mass. 
While using different approaches, several studies predicted a small core mass between 0.2 - 5 M$_{\oplus}$ for giant protoplanets. 
Already in \cite{P86} it was shown that the maximum core mass when considering accretion of $100$ km-sized planetesimals is between $1$ and $3$ M$_\oplus$.
\cite{Brouwers2017} investigated the growth of the core  with accretion of pebbles/planetesimal at early stages. 
It was found that pebble accretion leads to a core with a maximum mass between 0.1 - 0.6  M$_{\oplus}$ depending on the pebbles' composition. 
For the case of 1 km-sized rocky planetesimals a maximum core mass between  $0.2 - 1.2$ M$_{\oplus}$ was derived. It was shown that the predicted core mass depends on the assumed material strength of rock and its effect on  the planetesimal's ablation/fragmentation. 
\cite{Alibert17} performed a similar study, investigating the envelope's mass required to disrupt $10$ cm-sized pebbles. 
It was found that an envelope with a mass of $0.001$ M$_{\oplus}$ is sufficient to destroy the pebbles. The mass of the core was found to be between $0.5$ and $3$ M$_{\oplus}$.
\cite{Lozovsky17} investigated the distribution of heavy elements in proto-Jupiter accounting for different solid surface densities and planetesimal sizes. A maximum core mass of 2-3 M$_{\oplus}$ was found with the rest of the heavy elements having a gradual distribution throughout the planet. It was also shown that further settling of the heavy elements is negligible. 
All of the studies mentioned above support the concept of a small core for giant planets. However, it should be noticed that similar studies by \cite{Mordasini06} and \cite{Baraffe06} derive core masses of $\sim$ $5$-$6$ M$_\oplus$. Possible reasons for the higher inferred core masses could be different treatments of fragmentation, and different assumed material strengths and $C_h$ values (see discussions below).

%It was found that the core in at the traditional way (of a pure-heavy element central region) consists of  up to $5$ M$_{\oplus}$ with the rest of the heavy elements having a gradual distribution throughout the planet. 
%Finally, \cite{Pinhas2016} investigated the efficiency of planetesimal ablation in current-state Jupiter. It was found that most of the planetesimals are completely dissolved in the envelope for a wide range 
%of planetesimal sizes, compositions, and velocities. In this case, only large iron planetesimal can reach the core. The density and temperature profiles of Jupiter today, however, are very different from the ones during and shortly after its formation.   
\par

A fundamental aspect in predicting the heavy-element distribution in proto-Jupiter (and giant protoplanets in general) is linked to the interaction of the solids (which can be pebbles or planetesimals) with the planetary envelope. 
The fate of the heavy
elements is uncertain and depends on the physical properties of the
accreted planetesimals and of the gaseous envelope. 
In addition, the distribution could depend on the  treatment of planetesimal fragmentation and ablation. 
%In this study, however, planetesimal fragmentation was treated in a very simple manner 
%{\bf fix  that is the dominant process for the mass deposition of large planetesimal and they did not analyzed how the results are sensitive to different ablation parameter.}
In this study, we explore the interaction of the heavy elements
with the gaseous envelope accounting for the ablation and fragmentation of planetesimals and determine the heavy-element distribution within the planet. 
We also investigate the dependence of the heavy-element distribution and inferred core mass  on the planetesimals' properties and the treatment of the planetesimal-envelope interaction.  %We expand previous methods \citep{Podolak88,Lozovsky17} to include different models for planetesimal fragmentation. 
%We show the importance of fragmentation, that is the dominant mechanism for mass deposition of large planetesial ($>$ 1 km). 
%We discuss the importance of fragmentation for large ($>$ 1 km) planetesimals and show that in all the cases a pure heavy-element core reaches a mass up to $2 M_{\oplus}$, with the exact value depending on the model assumptions. 
Finally, we present a simple semi-analytical approach for deriving the heavy-element distribution. 
%{\bf this is not good!}
 %\st{In this paper we investigate how the predicted distribution of heavy elements in within giant protoplanets depends on the treatment of the planetesimal-envelope interaction and the various model assumptions.} 

\section{Methods}

The interaction of a planetesimal with the planetary envelope is simulated following the approach of \cite{Podolak88} where  at each step of the two dimensional trajectory we compute the planetesimal's motion in response to gas drag and gravitational forces (assuming a 2-body interaction).  The effects of planetesimal heating and ablation as the planetesimal passes through the envelope are also included. We also consider planetesimal fragmentation which is set to occur when the pressure gradient of the surrounding gas across the planetesimal exceeds the material strength and the planetesimal is small enough that self-gravity cannot counteract the disruptive effect of the pressure gradient, and is given by \citep{P86}: 

\begin{equation}
\label{breakup}
	P={\frac{1}{2}}\rho_{gas}v^2\ge S, % ;\,\,\,\,\,\,\,\,\,\,
%	r_p<r_{dyn} = \sqrt{\frac{5 v^2 \rho_{gas}}{8 \pi G \rho_p^2}}
	\label{breakupcondition}
\end{equation}
\begin{equation}
\label{breakup}
%	P=\bf{\frac{1}{2}}\rho_{gas}v^2\ge S ;\,\,\,\,\,\,\,\,\,\,
	r_p<r_{dyn} = \sqrt{\frac{5 v^2 \rho_{gas}}{8 \pi G \rho_p^2}},
	\label{breakupcondition}
\end{equation}
where $P$ is the pressure, $v$ is the planetesimal's velocity, $\rho_{gas}$ is the envelope's density, $S$ is the  compressive material's strength depending on the planetesimal's composition, 
 $\rho_p$ and $r_p$ are the planetesimal's density and radius, respectively.\footnote{Note that Equation 1 is sometimes written without the factor of $1/2$ \citep{Zahnle92,Hills93}, independently to whether the gas pressure is assumed to act only on the planetesimals' front or on its entire surface.}
We have implemented several improvements to the new computation. First, we include an adaptive step size control to the 4th order Runge-Kutta method that is used to solve the equation of motion.  
Instead of evaluating the equation of motion once per time step, we do it three times; once as a full step, and then, independently, as two half steps until convergence is found. 
Second, we use an improved model for the planetesimal's fragmentation. In \cite{Podolak88} when a planetesimal fragments it was assumed that the entire planetesimal mass is deposited in that layer,  %it was occupying. 
while we continue to follow the planetesimal's fragments considering different fragmentation models (see section \ref{breakupsection}). Further details on the atmosphere-planetesimal interaction are presented in the Appendix. 
\par

The equations describing the motion and mass loss of a planetesimal in the planetary envelope are:
\begin{equation}
 m_{pl}\frac{d\vec{v}}{dt}=-\Gamma A \rho_{gas} \left|v\right| \vec{v} -G\frac{M_{p}(r)m_{pl}}{r^3}\vec{r},	
\label{EqofMotion}
\end{equation}
\begin{equation}
\frac{dm_{pl}}{dt}=-\frac{A}{Q}\bigg(\frac{1}{2}C_h\rho_{gas} \left| v\right|^3+\epsilon \sigma T_{a}^4\bigg).
\label{massablation}
\end{equation} 
The first is the equation of motion  in 2D, where $m_{pl}$ is the planetesimal's mass, $\Gamma$ is the drag coefficient (calculated as in \cite{Podolak88}), $A$ is the planetesimal's  surface, $r$ is the distance from the protoplanet's center and $M_p(r)$  is the planet's mass inside $r$. 
The second equation describes the planetesimal's ablation where $T_a$ is the atmospheric background temperature and $\sigma$ is the Stefan-Boltzmann constant.
There are two sources for ablation: the radiation from the surrounding atmosphere and gas drag. % the goes to ablate the planetesimal. 
For simplicity, the atmosphere is assumed to behave as a  gray body, $\epsilon$ is the  emissivity which is a product of the emissivity of the atmospheric gas and the absorption's coefficient of the impactor.  The value of $\epsilon$ is not well-determined and depends on the local density, pressure, temperature, and composition of the atmosphere. Therefore, we assume different $\epsilon$ values, and investigate their impact on the results. % \st{to the assumed value}. 
$Q$ is the latent heat caused upon vaporization, and $C_h$ is the heat transfer coefficient. 
$C_h$ is the fraction of the relative kinetic energy transferred to the planetesimal and  
%$\frac{1}{2}S\rho v^3$ of the oncoming stream of molecules is expended on the ablation of the mass $dM$ in time $dt$ and
 its value can range between zero and one.   
Apart from the energy associated with ablation, a fraction of the energy heats up the planetesimal itself, and the rest of the energy is converted into radiation that ionizes  the atoms and molecules of both the planetesimal and the atmosphere. 
If fragmentation is considered, the portion of energy leading to fragmentation (i.e. breaking the mechanical bonds between particles) must be included.  
Essentially, the division in energy to the different processes is embedded in the $C_h$ value. 
\par

An accurate determination for $C_h$ requires complex 3D radiation-hydrodynamic (RHD) and computational fluid dynamics (CFD) simulations (\citealt{Makinde13}, \citealt{Pletcher12}, \citealt{Nijemeisland04}). 
An upper limit to $C_h$ is given by $C_h=\Gamma/2$ \citep{Allen1962}. 
Different studies assume different values for $C_h$. A value of $0.1$ was assumed in \cite{Podolak88} and \cite{Inaba03}, while \cite{Pinhas2016} used $C_h=10^{-2}$. 
\cite{Mordasini2015} assumed values between $10^{-3}$ and $10^{-5}$ following the suggestion of \cite{Svetsov1995}. % and \cite{Apstein1989}. 
The low values for $C_h$ are derived from simulations of the entry of comet Shoemaker-Levy9  to Jupiter's atmosphere, while the higher values are typically inferred for objects hitting the Earth's atmosphere although there were also applied to current-state Jupiter \citep{Pinhas2016}. 
Naively one would think that for our purpose a Jupiter-like atmosphere is more relevant, but Shoemaker-Levy9 might not represent the standard case, and in addition, the actual value of $C_h$ can significantly change as it passes throughout the atmosphere and loses mass. 
As a result, it is unclear which value is most appropriate, and in this work we use various values for $C_h$, and explore how they affect the heavy-element distribution in the protoplanetary envelope. 
\par

We set the solids to be represented by planetesimals, and consider three different sizes of  %An analysis of the interaction with pebbles is shown by \cite{Brouwers2017} \st{ some information with planetesimal accretion} (\cite{Mordasini2015}). 
100 km, 1 km, and 10 m,  as well as different compositions: rock, water, and a mixture of rock+water.  
%\textit{You already say this thing down so I would remove it here} \st{The rate at which the core grows is equal to the total amount of solid accreted by the planet, that in our model is $\dot{M}_Z=10^{-6}M_{\oplus}/yr$}. 
Following \citet{P96}, a planetesimal composed of a mixture of water and silicates (rock+water) is assumed to be 50\%-rock  and  50\%-water by mass with the rocky material  being  embedded in a matrix of water ice. % In this case, the ice acted as the ��glue�� that held the planetesimal together. and that the rocks surrounded by water ice. 
When the ice around the rock is vaporised the rock in this layer is also assumed to be released into the envelope as ablated material, keeping the planetesimal's composition unchanged. 
All planetesimals are assumed to have an initial velocity of $1$ km/s directed on the $x$-axis.  
% the rock grain will no longer be bound to the planetesimal and will be lost. {\bf this is still very badly written!!!! As a result the density and the composition of the planetesimal are constant.}
\par

\begin{figure}
	\centering
		\includegraphics[width=0.350\linewidth]{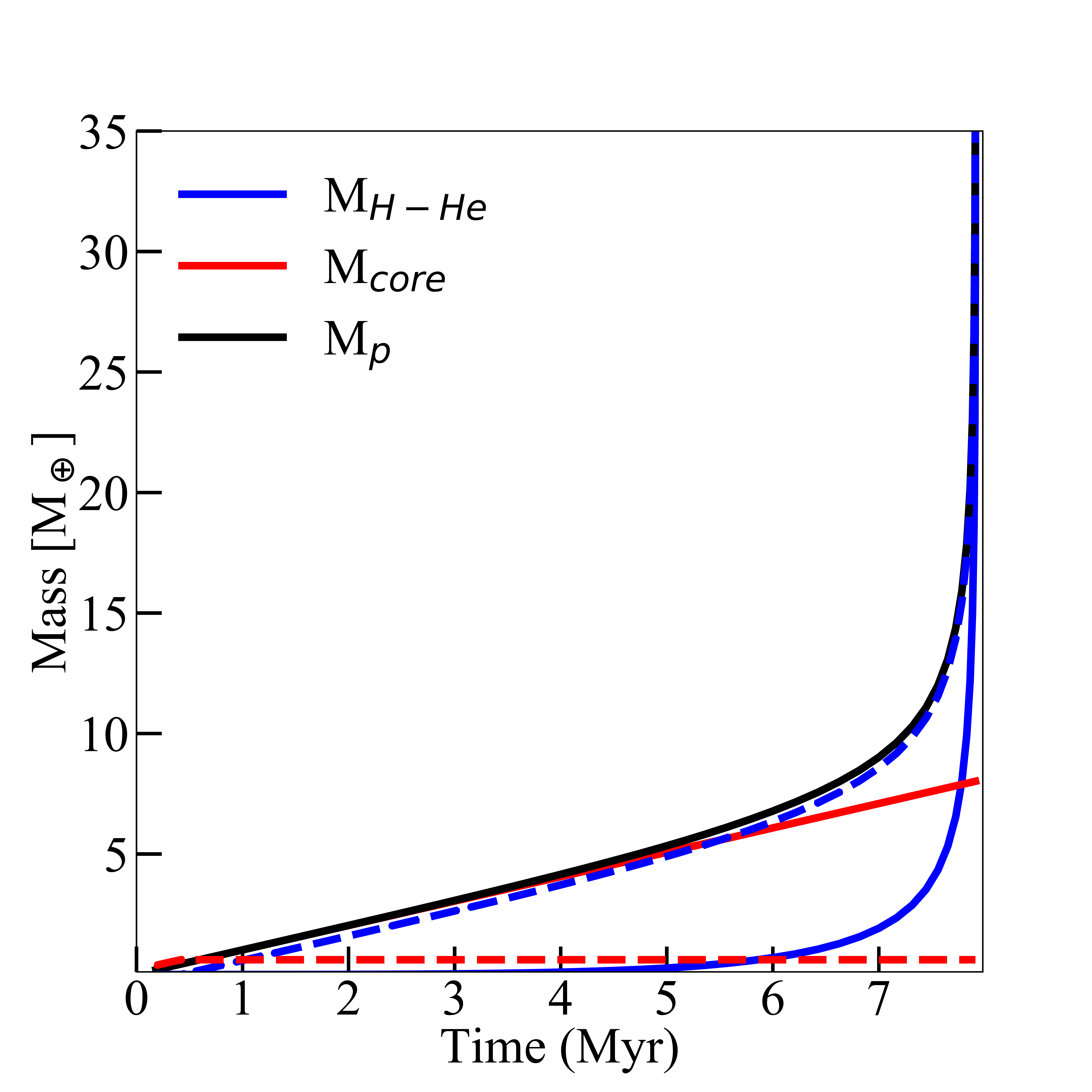}
			\vskip -10pt
	\caption{The envelope mass (blue), the core mass (red) and the total planetary mass (black) as a function of time. The solid accretion rate is set to be   $10^{-6}$ M$_{\oplus}$/yr. 
	The  dashed line corresponds to a case in which the heavy elements are deposited in the envelope (enriched) while the case of the solid line correspond to a non-enriched case in which all the heavy elements are assumed to reach the center (pure H-He envelope). The planetesimals are assumed to be  composed of water with sizes of 10 m. More details about the two cases can be found in \cite{Venturini16}.}
	\label{AtmoMass}
\end{figure}

For the background atmospheric models and planetary growth we use a standard core accretion planet formation simulation kindly provided by J.~Venturini. The model corresponds to Jupiter's formation at 5.2 AU with solid surface density of 10 g/cm$^{-2}$ and solid accretion rate of $10^{-6}$ M$_{\oplus}$/yr. 
The dust and gas opacities are respectively given by \cite{Mordasini14} and \cite{Freedman14}. The EOS for the H-He envelope is taken from \cite{Saumon1995} (see \cite{Venturini16} and Figure \ref{AtmoMass} for details). 
The planetary growth is computed assuming that all the accreted heavy element mass goes to the core and the envelope's composition is a mixture of hydrogen and helium in proto-solar ratio. 
In this setup, the formation timescale for Jupiter is $\sim 8\times 10^6$ years. Figure \ref{AtmoMass} shows the modelled planetary growth for the standard case where all solids goes to the core, and for the case where envelope enrichment (planetesimal ablation) is considered (see next section for details).

\subsection{Capture Radius and Inferred Core Mass}
\begin{figure}

	\centering
	\includegraphics[width=0.3\linewidth]{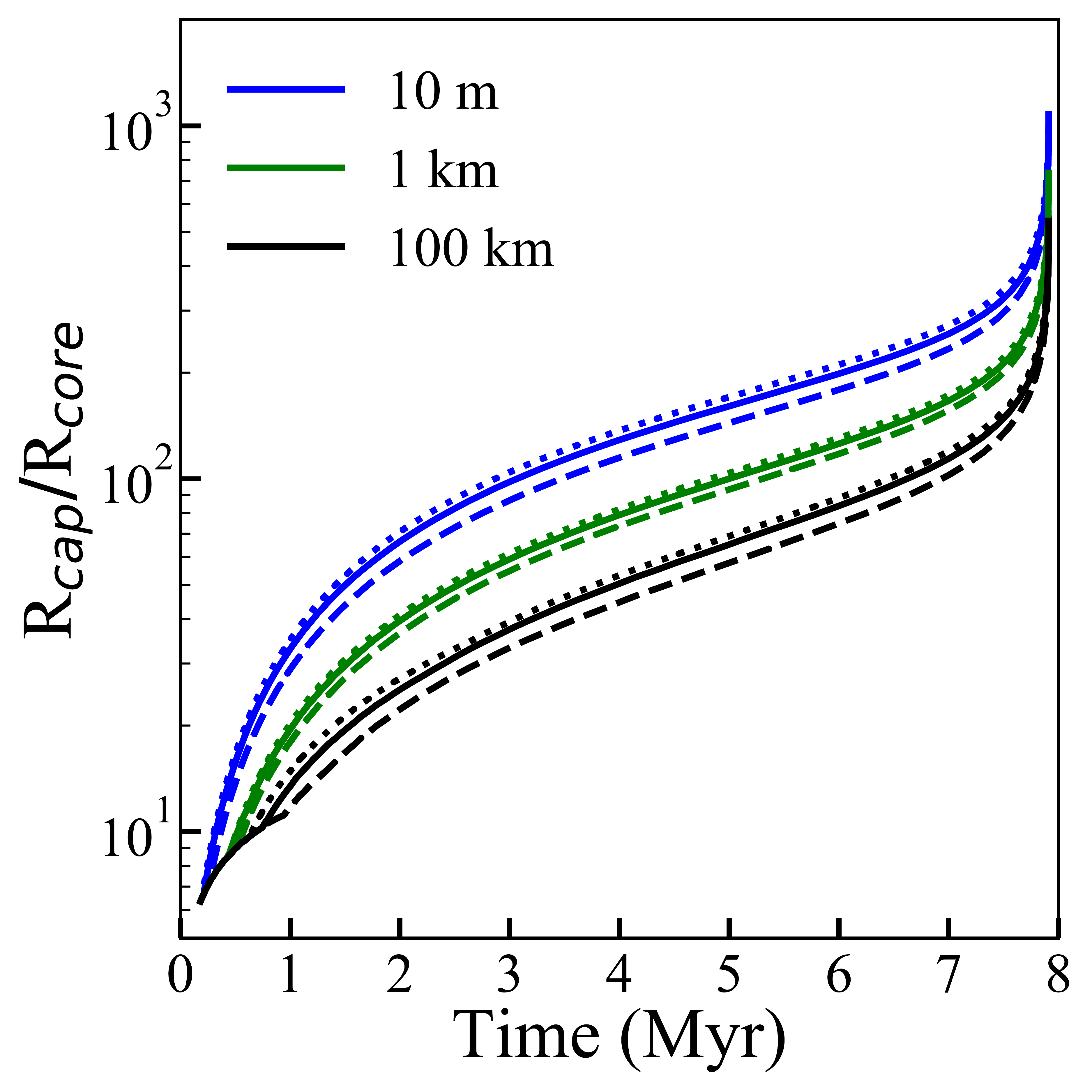}
		\includegraphics[width=0.3\linewidth]{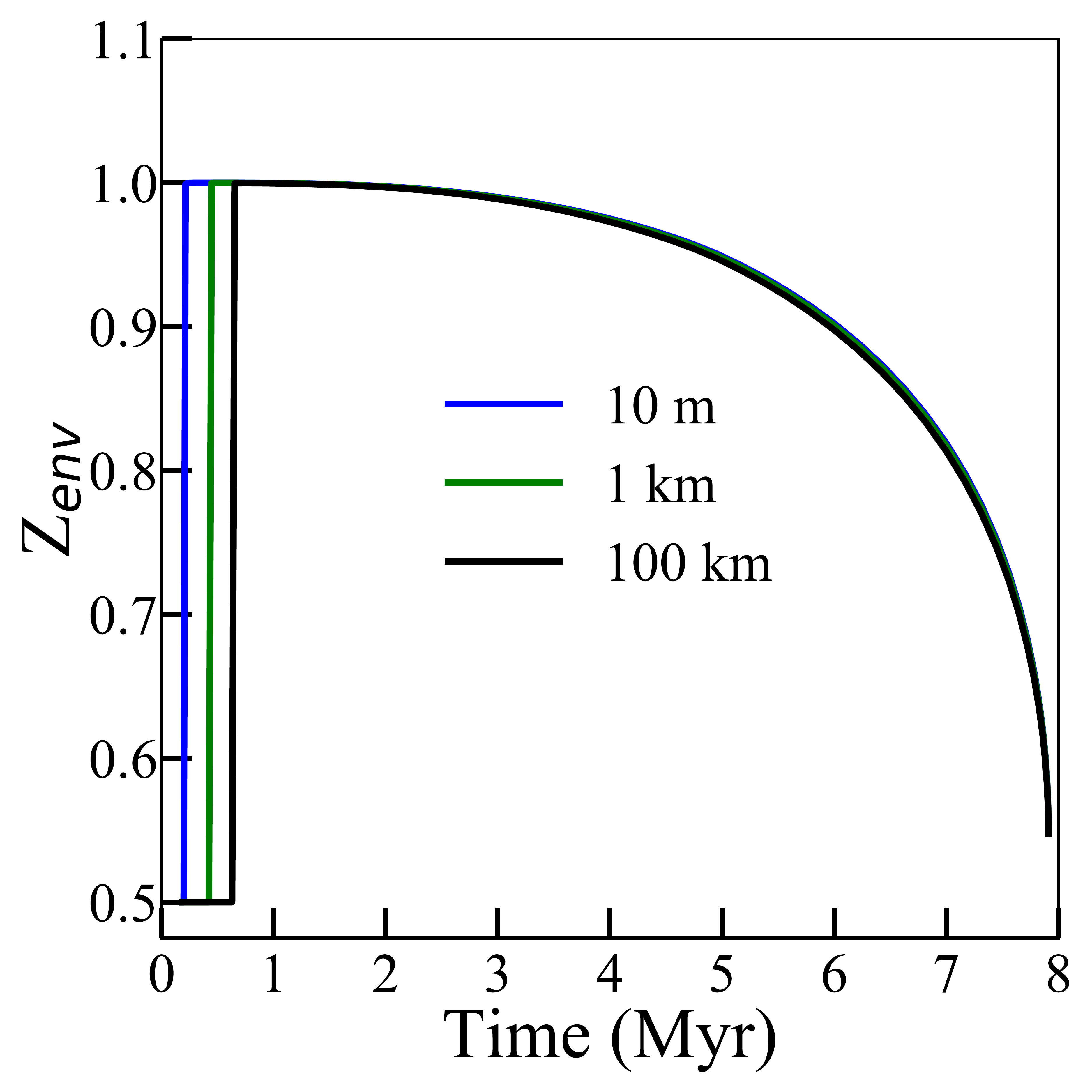}
	\includegraphics[width=0.3\linewidth]{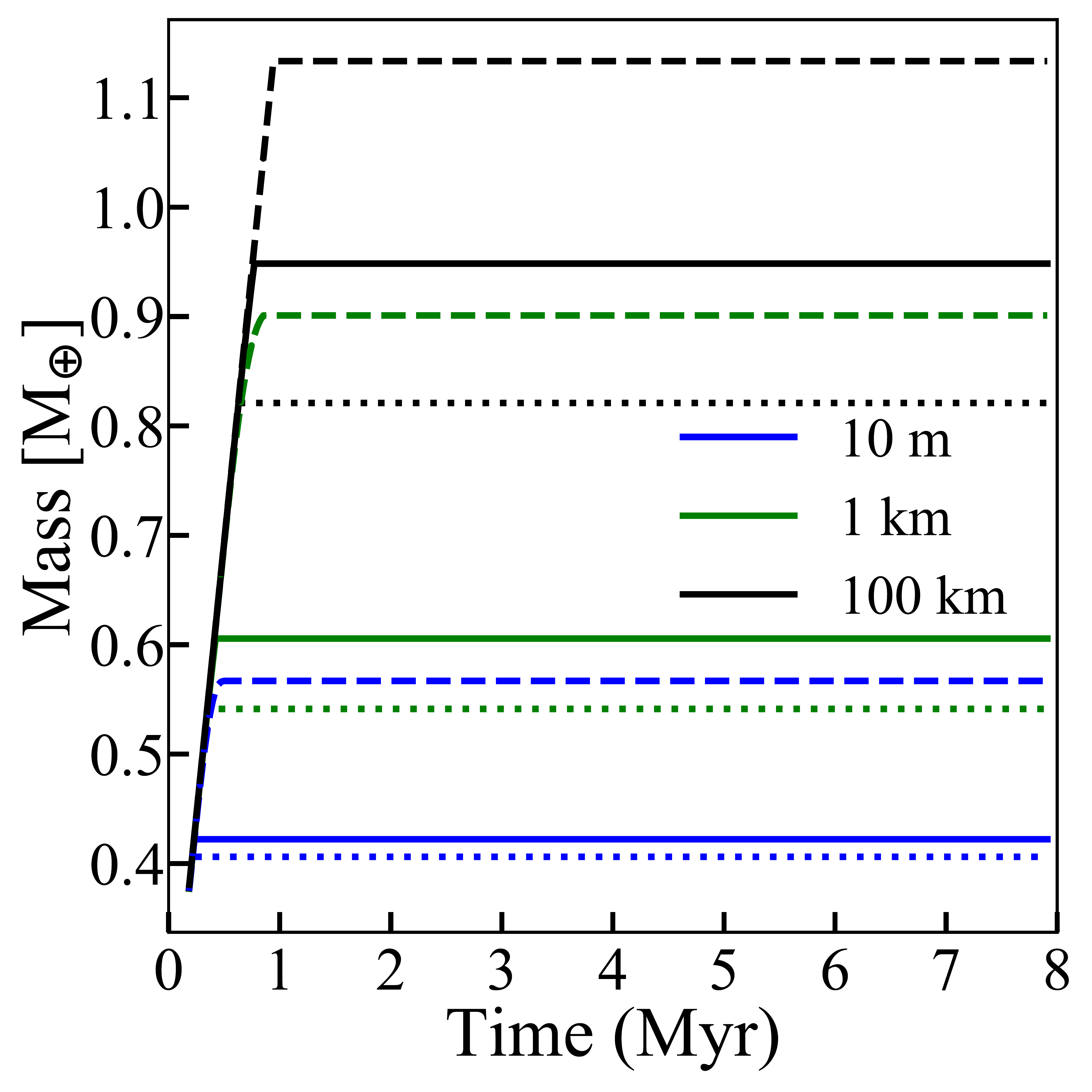}
		\includegraphics[width=0.3\linewidth]{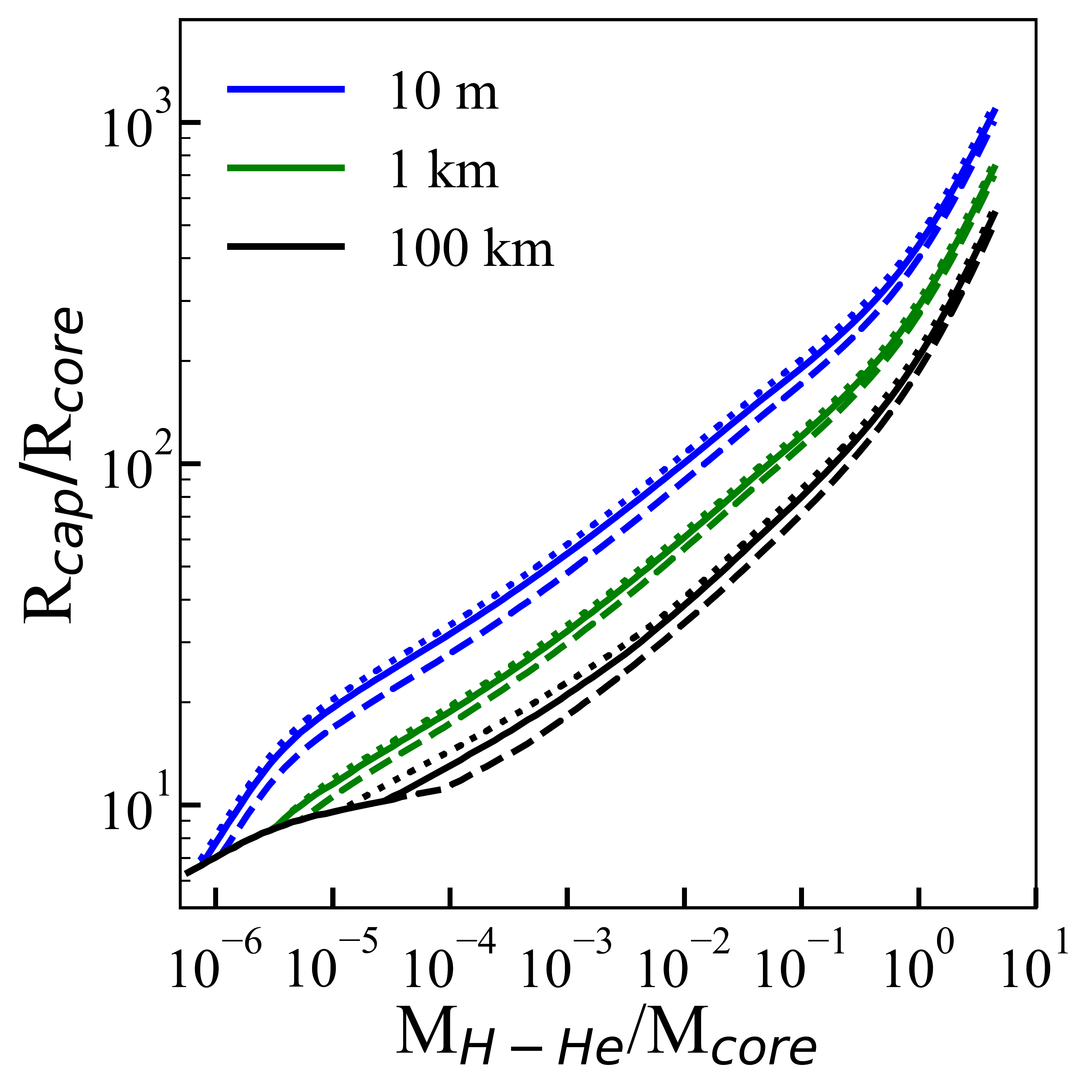}
		\includegraphics[width=0.3\linewidth]{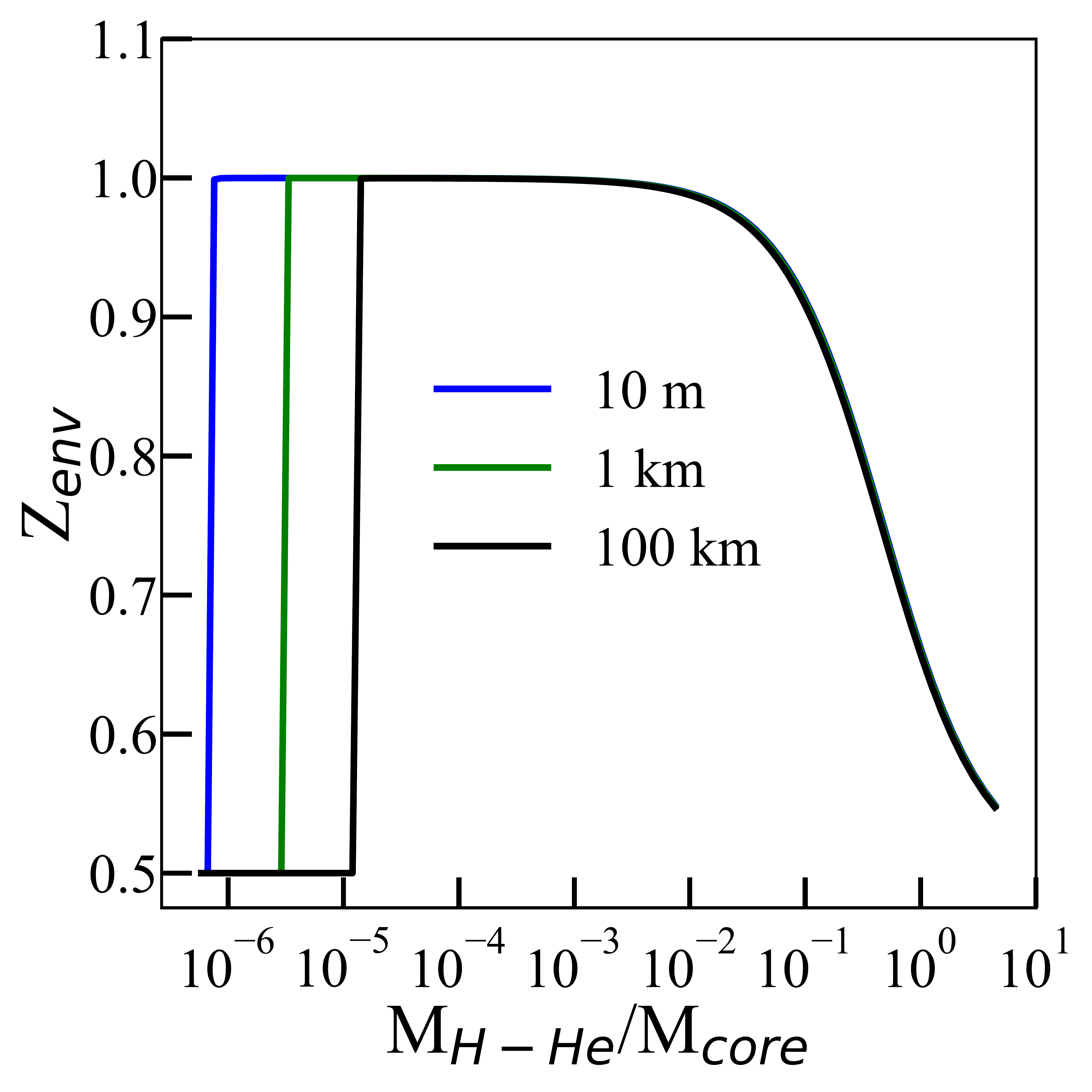}
	\includegraphics[width=0.3\linewidth]{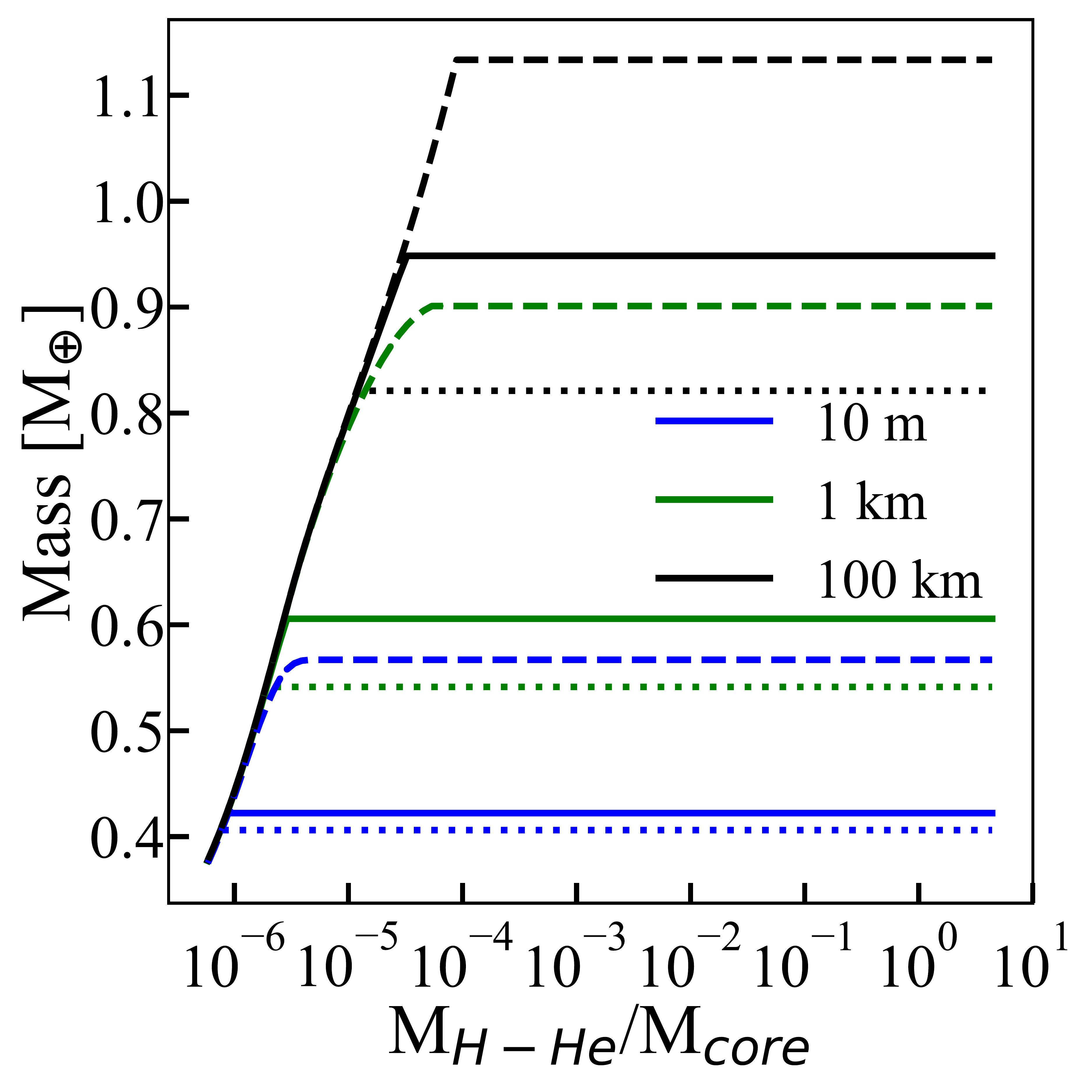}
	\caption{{\bf Left:} Capture radius over core radius $R_{cap}/R_{core}$ for different planetesimal sizes. The dotted, solid, and dashed curves correspond to water, ice+rock, and rocky planetesimals, respectively. {\bf Middle:}  The envelope's metallicity $Z_{env}$ vs.~time for three different planetesimal sizes. The planetesimal composition is assumed to be water+rock. {\bf Right:} Inferred core mass for the different cases. The dotted, solid, and dashed curves correspond to water, ice+rock, and rocky planetesimals, respectively. The top and bottom panels show the above properties vs.~time and $M_{H-He}/M_{core}$, respectively. }
		\label{Fig2}
\end{figure}

At early stages, the planetesimals go through the envelope and reach the core, although some of their mass is deposited in the atmosphere. 
As time progresses, planetesimals no longer reach the center, and instead, their mass is deposited in the envelope, leading to envelope enrichment \citep{Podolak88,Mordasini06,Fortney13}. 
%
% % (\citep{Pinhas2016,Ormel17,Lozovsky17}). \textbf{here}
%We find that when planetesimal ablation and fragmentation are included, the core mass is found to be small, and after a short time, for all the planetesimal sizes and compositions we consider, its mass reaches a maximum value in agreement with previous studies (\citealt{Lozovsky17}, \citealt{Brouwers2017}).  
%%It is found that for all  planetesimal sizes and compositions the core reaches a maximum mass. %I.e., from a given time on (depending on the exact size/composition), no planetesimals reach the core. 
%%The core reaches the maximum core mass quite fast at a time of $\sim$ 1-2 Million of years  obtaining a final small core value.}
%%At early stages, during the first 2 Million of years, H-He is accreted very slowly and the envelope mass is small. However, once solids stop reaching the core, the mass fraction of heavies in the envelope increases with time, and can be as large as a factor of 1000 in comparison to  the H-He mass. Of  course in this situation even if the heavies are in a vapour phase most of the planet is composed of heavy-elements, and this configuration can be viewed as a diluted/fuzzy core \citep{Lozovsky17,Ravit17}. 
%During these formation phases there is no sharp boundary between the core and the envelope in terms of composition and the core region is not well defined. 
%First, we calculate the capture radius for different planetesimal sizes and compositions (see Appendix for details).  
%The results are shown in the left panel of  Figure \ref{Fig2}.  % all possible cases are compared.
The left panel of  Figure \ref{Fig2} shows the capture radius $R_{cap}$ for different assumed planetesimal sizes and compositions. 
 This plot demonstrates the importance  of gas drag in determining the planet's capture radius (and therefore the solid accretion rate). 
%It is clear that if we neglect gas drag and ablation the capture radius is independent of the planetesimal's physical properties. 
The importance of accounting for the ablated heavy-element mass in the atmosphere in planet formation models is reflected by the difference between the three curves. 
Small planetesimals have larger capture radii and are captured more easily. 
%If we think that the distribution of planetesimal is uniform, not dependent on their sizes, so that in a sphere of radius $R$ there is the same amount of planetesimal of every size, the plot show that the accretion rate of smaller planetesimal should  be taken as $6-10$ greater than the bigger planetesimal one (mass that is in a sphere grows as $R^3$).
As expected, water planetesimals are captured more easily than rocky ones. %, and smaller planetesimal are captured more easily than large ones.  
The figure also shows that the planetesimal size, rather than composition, is the dominating parameter in determining the capture radius. 
The capture radius is determined by searching for the largest value of the impact parameter ($b_{crit}$)  for which the planetesimal is captured (e.g., \citealt{Ravit06}). 
Further details on the capture radius are given in the Appendix.
%A planetesimal is captured when it reaches the planetary core or is vaporized by the envelope.}

The middle panel of Figure \ref{Fig2} shows the ratio between the heavy-element mass in the envelope ($M_{z,env}$) and the H-He mass. 
While the solid accretion rate is constant and equals to 10$^{-6}$ M$_{\oplus}$/yr the accretion rate of H-He increases with time. As more H-He is accreted by the growing planet, $M_{z,env}$/$M_{H-He}$ decreases and finally the gas accretion rate exceeds that of the solids and the envelope's metallicity decreases significantly. 
It is interesting to note that during the early stages, when the gaseous mass is still small, the peak of the heavy-element mass ratio ($M_{z,env}$/$M_{H-He}$) occurs at different times, and the exact value of $M_{z,env}$/$M_{H-He}$ depends on the assumed planetesimal size. Nevertheless, in all the cases, since the final envelope composition is dominated by the gas accretion, once the planet reaches Jupiter's mass the envelope's metallicity is found to be $Z \approx 0.01$. 
The exact value, however can change depending on whether planetesimals are expected to be accreted at later stages (see \citealt{Helled14}). 
\par

As expected, ablation of planetesimals significantly changes the atmospheric mass, and larger planetesimals tend to reach the center and leading to less significant enrichment of the atmosphere. If heavy-element ablation is neglected the atmospheric mass is significantly smaller, which affects the subsequent planetary  growth and the ablation of planetesimals at successive time step as well as the evolution of the atmosphere itself due to change in opacity and the envelope's composition (equation of state). So far, this effect has only been considered by \citealt{Venturini16}, and clearly, this effect is significant.% behind less mass on the atmosphere. 
\par

The inferred core mass for the different cases is presented in the right panel of Figure \ref{Fig2}. 
%{\bf fix - there are only two stages - better think about the results and not about the phases...}\\
%\textbf{The plot shows that at the very beginning of the planetary growth, the atmosphere is very thin and the accreted solids can reach the core, 
%which grows linearly in time. After a certain period the density of the atmosphere is high enough to ablate or fragments the solids. 
We find that core growth occurs only at very early times (less than 1 Myrs) and that the core mass is rather small (less than 1.5 M$_{\oplus}$). 
After that point the envelope is dense enough to ablate/fragment the accreted solids, and the heavy elements stop reaching the center keeping the core mass constant. %and the core mass remains constant. 
The exact time at which the core stops growing and the final core mass depend on the properties of the envelope and the size and composition of the accreted planetesimals.  However, in all cases the core stops growing early and its mass remains small. % however, 
This confirms that the core accretion 
scenario can naturally lead to the formation of small cores, unless the accreted planetesimals are extremely large ($\gg$ 100 km). 
We find that when planetesimal ablation and fragmentation are included, the core mass is found to be small, and after a short time, for all the planetesimal sizes and compositions we consider, its mass reaches a maximum value in agreement with previous studies (\citealt{Lozovsky17}, \citealt{Brouwers2017}). Indeed, it was found by \cite{P86} that, except for impactors with sizes larger than  $1000$ km, the core mass stops increasing when it reaches a mass between $1$ and $2.8$ M$_{\oplus}$.
%It is found that for all  planetesimal sizes and compositions the core reaches a maximum mass. %I.e., from a given time on (depending on the exact size/composition), no planetesimals reach the core. 
%The core reaches the maximum core mass quite fast at a time of $\sim$ 1-2 Million of years  obtaining a final small core value.}
%At early stages, during the first 2 Million of years, H-He is accreted very slowly and the envelope mass is small. However, once solids stop reaching the core, the mass fraction of heavies in the envelope increases with time, and can be as large as a factor of 1000 in comparison to  the H-He mass. Of  course in this situation even if the heavies are in a vapour phase most of the planet is composed of heavy-elements, and this configuration can be viewed as a diluted/fuzzy core \citep{Lozovsky17,Ravit17}. 
It should be noted, however, that during these formation phases there is no sharp boundary between the core and the envelope in terms of composition and the core region is not well defined (e.g., \citealt{Ravit17}). 
%First, we calculate the capture radius for different planetesimal sizes and compositions (see Appendix for details).  
%{\it 
%The plot suggests that the growth of the core mass can be divided into three main phases: (1) At the very beginning of the planetary growth, the atmosphere is very thin and the accreted solids can reach the core. (2) the density of the atmosphere is high enough to ablate small solids due to gas drag (for our specific model at 0.5 Myr). Large planetesimal can still reach the core without being significantly ablated. (3) the density of the atmosphere is high enough to break-up the planetesimal ($1-1.5$ Myr). Solids can no longer reach the core directly. 
%In this phases, small solids enrich the outer layers of the atmosphere, while large planetesimal the inner part closer to the planetary center. 
%This result is in agreement with previous studies \citep{Iaroslav07,Lozovsky17,Ormel17} and confirms that the core accretion mechanism naturally leads the formation of small cores unless the accreted solids are extremely large $\gg$ 100 km. 
%}

\section{The dominant mechanism: ablation or fragmentation?}
\begin{figure}
	\centering
	\includegraphics[width=0.80\linewidth]{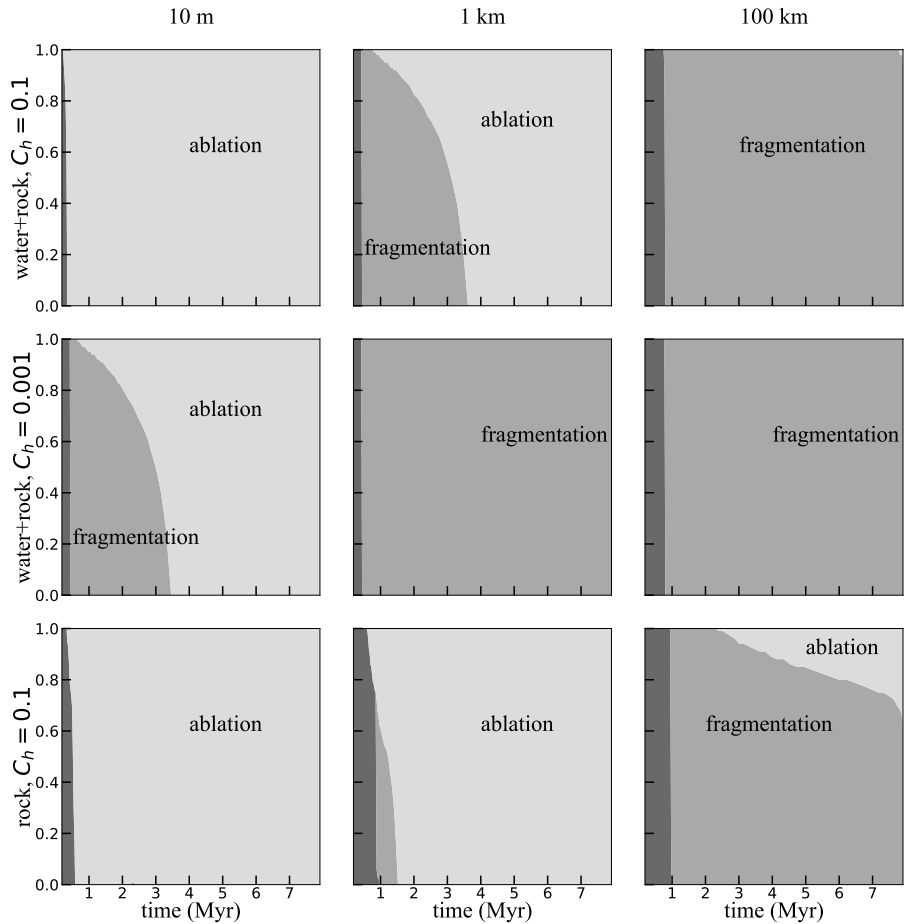}
	%\vskip{-0.2cm}
	\vskip -25pt
	\caption{The fate of infalling planetesimals vs.~time for different planetesimal properties and $C_h$ values. The different gray scales (from dark to light) correspond to the cases of reaching the core, fragmentation, and ablation, respectively. 
	%\textbf{On the 
	The $x$ axis shows the time, ranging from $0$ to $7$ Myr, while the $y$ axis shows the impact parameter of the planetesimal in unit of $b_{crit}$.
	The first column corresponds to 10m-sized planetesimals, while the second and third columns are for 1km-sized and 100km-sized planetesimals, respectively. 
	The first row corresponds to a composition of water+rock, $C_h=0.1$, the second row is the same composition but with $C_h=0.001$, and the third row is for rocky planetesimals with $C_h=0.1$.}
	\label{PlanetesimalFaith}
\end{figure}

%\section{Faith of Planetesimal}
While the result that once the core mass reaches a value of $\sim 1-2$ M$_{\oplus}$ is robust,  the actual distribution of the heavy elements in the envelope depends on whether the dominated mechanism is ablation (Equation \ref{massablation}) or fragmentation (Equation \ref{breakup}). 
Fragmentation often dominates the mass deposition of large planetesimal  ($\geq$ 1 km) in the inner regions of the envelope, where both the planetesimal's velocity and the atmospheric density are high. 
Small solids are mostly ablated and are typically deposited at higher regions in the envelope. %do not reach high velocities and so the right hand term in \ref{breakupcondition} is always small. Fragmentation is important for planetesimal of size $1$ Km and more.
Figure \ref{PlanetesimalFaith} shows the dominating mechanism for different planetesimal sizes and composition as the protoplanet grows. 
%We also show  we can see the faith for different planetesimal models. The plot shows that three parameter are important to understand when planetesimal break-up
%\begin{itemize}
The larger the solids are, the more likely it is that they fragment since they are less affected by ablation \citep{Mordasini2015}. 
%The velocity will increase, and thus the term in \ref{breakupcondition} will be higher. Modification to material strength due to self gravity is not considered
The planetesimal's composition also plays an important role - water planetesimals have a lower material strength $\sigma$ comparison to rocky planetesimals, and can therefore fragment more easily. 

Finally, the choice of the $C_h$ value is also important - the lower $C_h$ is the more likely it is that fragmentation occurs. This is because for low $C_h$ values ablation is less significant and planetesimal can reach deeper regions within the envelope where the density is high enough to cause fragmentation. In the case of big bodies the value of $C_h$ is expected to vary inversely proportional with the atmospheric density reaching value of $\sim 10^{-4}$, as can be seen adopting the formula given in \cite{Melosh08}  
The prediction that the core mass remains small is insensitive to the dominating "deposition mechanism". In all cases after $\sim $1 Myr all the accreted solids are either ablated and fragmented and can no longer reach the core. 
\par 

\begin{figure}[h!]
\centering
\includegraphics[width=0.50\linewidth]{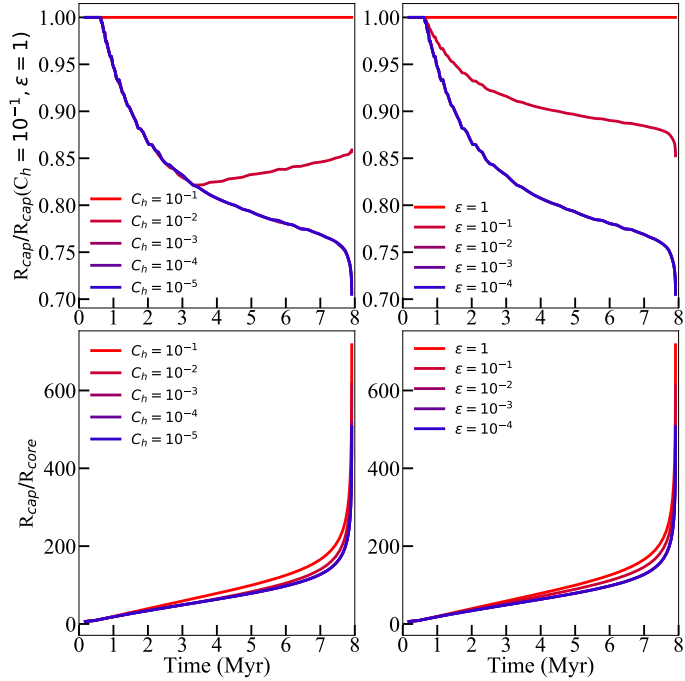}
\caption{ {\bf Top: }$R_{cap}/R_{core}$ for different $C_h$ and $\epsilon$ values. {\bf Bottom:} {$R_{cap}/R_{cap}(C_h=0.1,\epsilon = 1)$  for different $C_h$ and $\epsilon$ values.} The planetesimals are assumed to be  composed of water+rock and $1$ km in size.} %. The label $1$ in legend stand for a $C_h$ value equals to $0.1$, label $10$ for $0.1/10$, label $20$ stands for $=.1/20$ and so on}
\label{ChCaptureRadius}
\end{figure}

Figure \ref{ChCaptureRadius} shows the inferred capture radius for different assumed $C_h$ (left panel) and $\epsilon$ (right panel) values for 1 km-sized planetesimals composed of water+rock. 
%\st{The left panel shows the sensitivity of $R_{cap}$ to $C_h$.} 
As expected, larger $C_h$ and $\epsilon$ values lead to more efficient capturing and larger capture radii. 
Interestingly, a change of $C_h$ and $\epsilon$ by several orders of magnitude results in a difference of up to a factor of two in the capture radius. Changing $C_h$ and/or $\epsilon$ leads to very similar results.
%From one extreme ($0.1$) to the other $10^{-5}$ we see that in total the capture radius changes by a factor that is between $1$ and $2$. 
This confirms that $R_{cap}$ is insensitive to ablation as noted by \cite{Inaba03}.
%This suggests that $R_{cap}$, and therefore the solid accretion rate, is insensitive to the assumed values of $C_h$ and $\epsilon$. % value and is therefore well-determined.  
%Thus, in this plot, we have shown that the capture radius, or th feeding zone of the planet, can be thought of being independent on the $C_h$ value, that is a very unknown parameter in friction physics. 
In addition, we show that $R_{cap}$ is nearly unchanged for $C_h$/$\epsilon$ values between 0.1 and $10^{-3}$. Only extremely low values can slightly change $R_{cap}$, but even then, the change is insignificant. 
This is because at such a low value ablation becomes completely negligible. %, and the capture radius becomes a quantity that depends only on the geometry of the system, and thus independent on $C_h$. 
However, as we show in the next section, the $C_h$ value has an important role in determining the heavy-element distribution in the planetary envelope. 
\par

%It should be noted that $b_{crit}$ also depends on the $C_h$ value. % this plot we show also that, when you run the simulation to compute where planetesimal deposit their mass, you have to keep  in account also that $b_{crit}$ changes if you change $C_h$ 

\section{The Distribution of Heavy Elements}
\begin{figure}[h!]
	\centering
	\includegraphics[width=0.3275\linewidth]{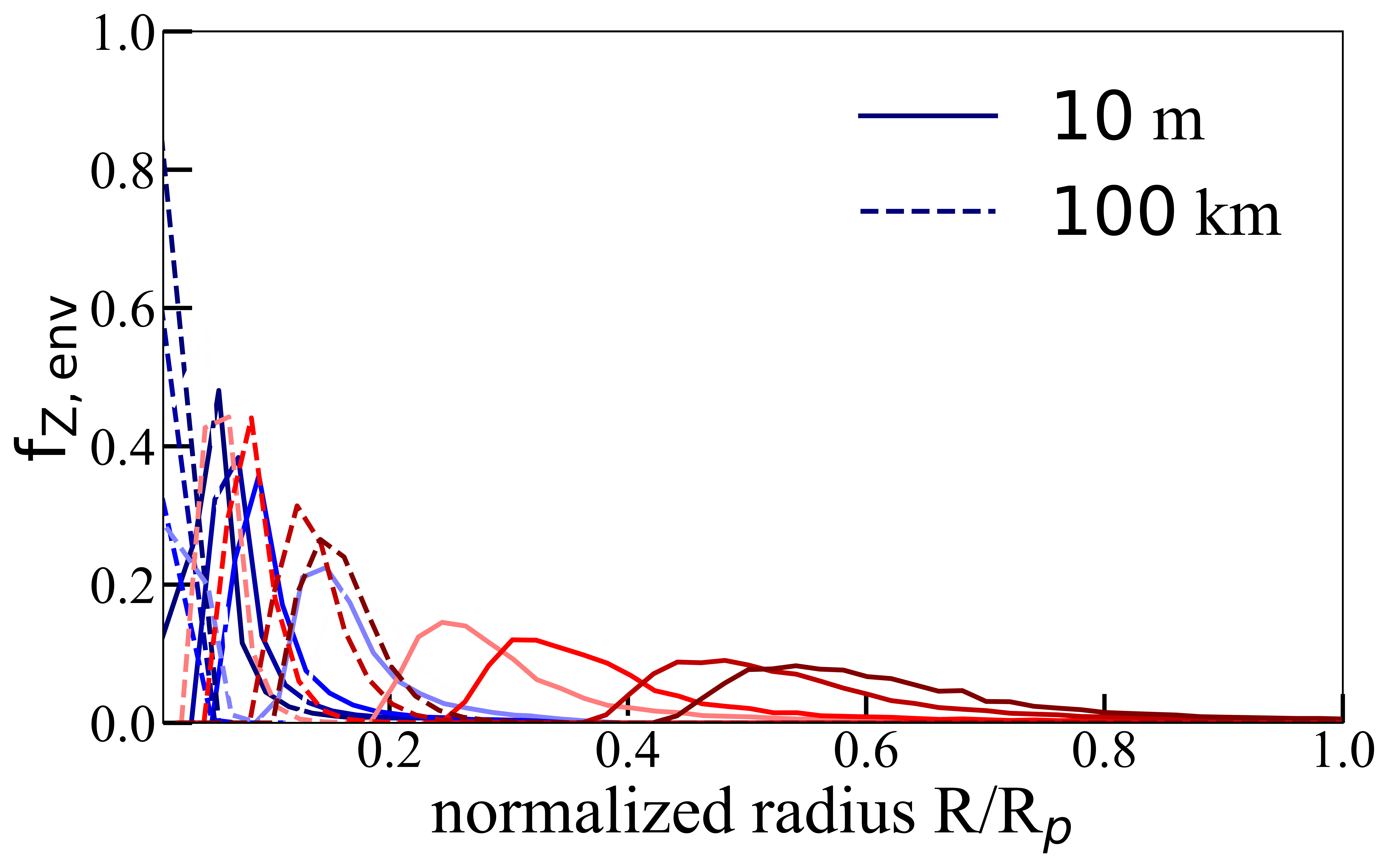}
	\includegraphics[width=.3275\textwidth]{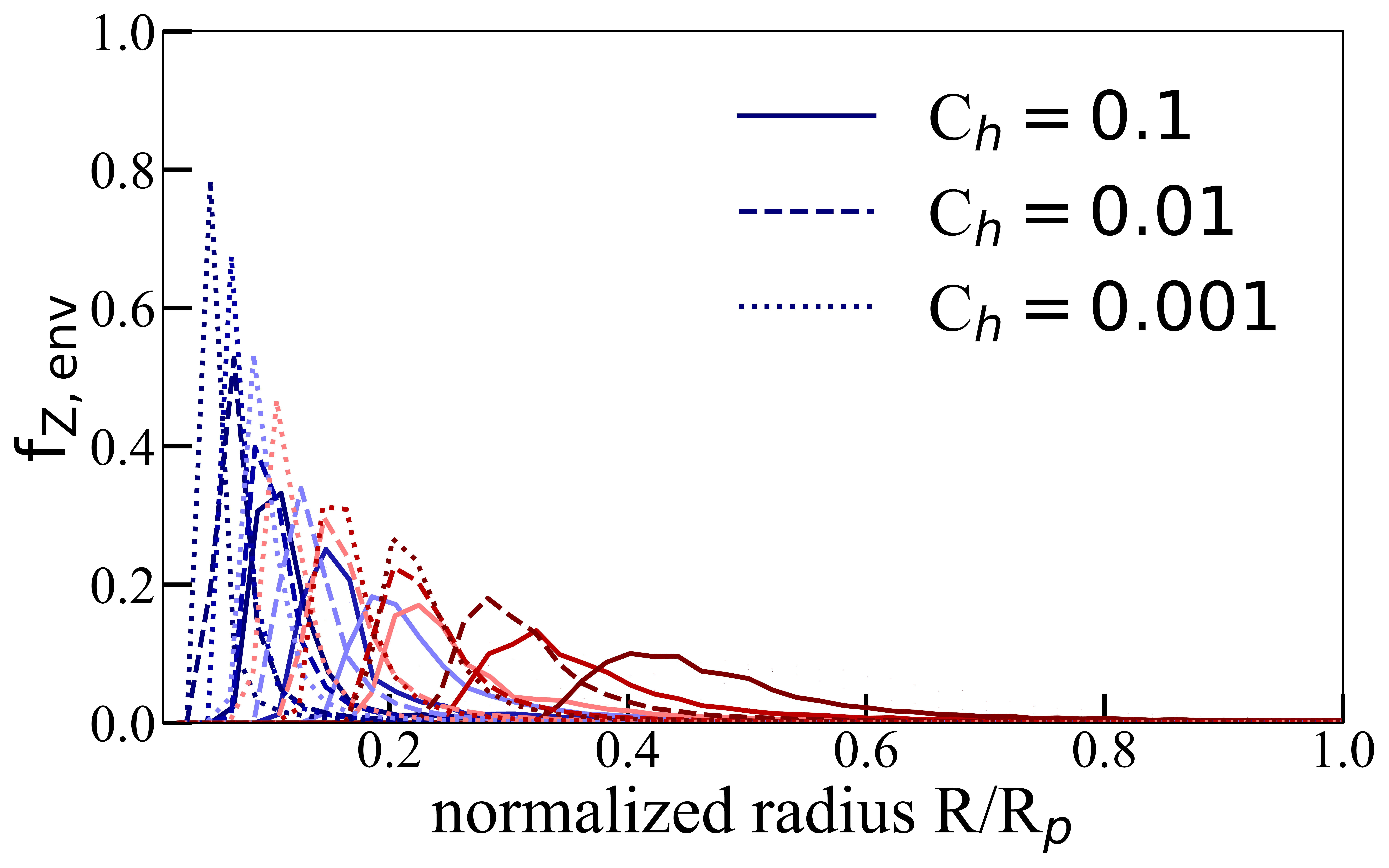}
	\includegraphics[width=.3275\textwidth]{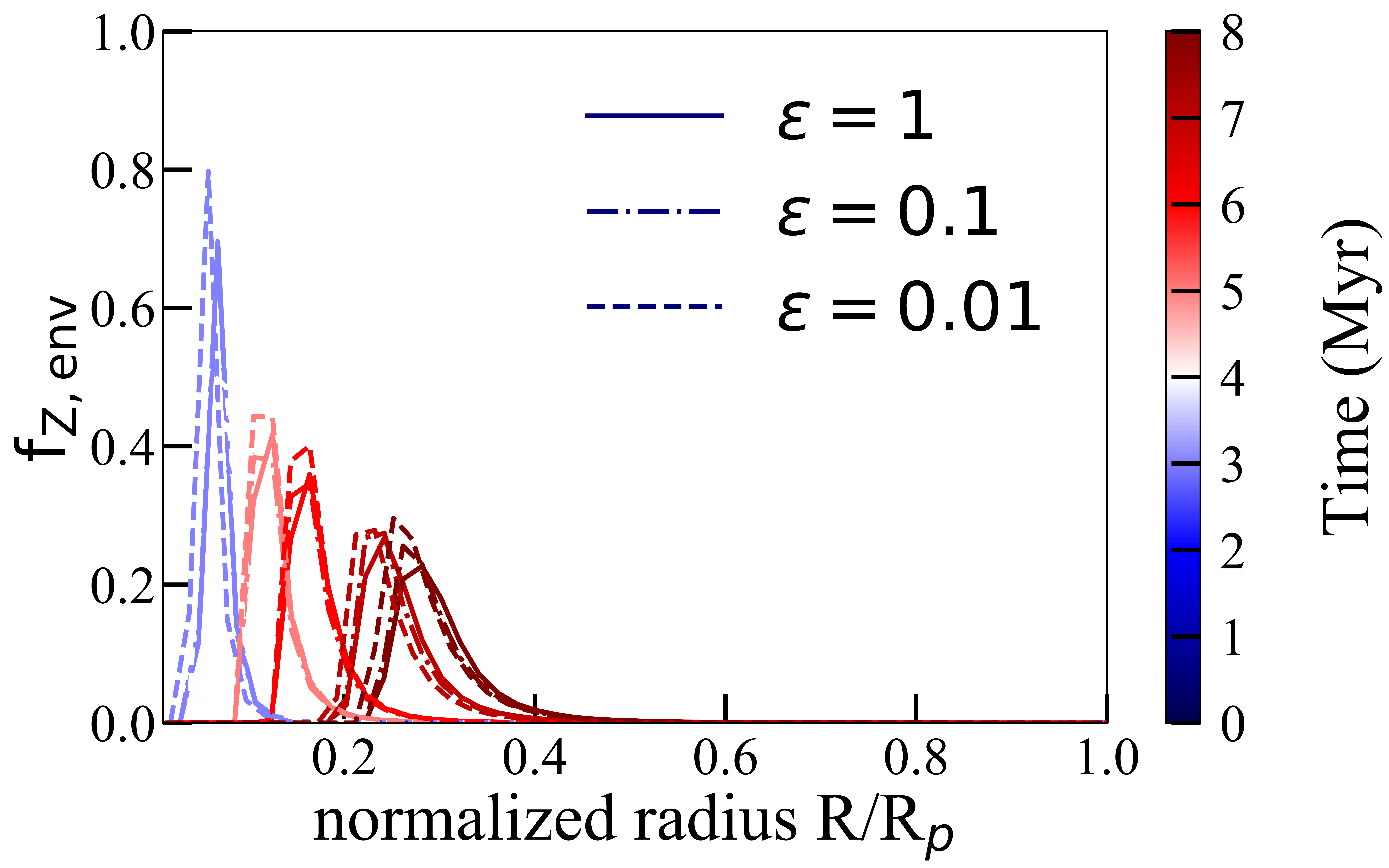}
	\caption{$f_{Z,env}$ vs. normalized radius at different times. The times are shown in the colour scale from red (10$^5$ years, early) to blue ($8\times 10^6$ years, late).  The plots are produced dividing the total radius of the planet in $50$ bins, each with a width of $0.02$ $R_{tot}$. These results are also presented in the Appendix where $f_{Z,env}$ at different times is shown in separate panels.
	\textbf{Left:} The solid and dashed lines correspond for planetesimals composed of a mixture of rock+ice with sizes of $10$ m and 100 km, respectively. {\bf Middle}: Rocky 10 m-sized planetesimals with $C_h=0.1$ (solid), $C_h=0.01$ (dotted) and $C_h=0.001$ (dashed-dotted). \textbf{Right}: Rocky 1 km-sized planetesimals with different $\epsilon$ values. The solid, dashed and dotted stands for $\epsilon$ of 1, $0.1$ and $0.01$, respectively.} 
	%\it{Here I would do maybe like this. I would remove this bottom plot with epsilon as the difference between the plot is small I would simply say the the heavies distribution are not super sentivie to the epsilon value, due to the fact in the inner layer the gas drag is the main source of ablation rather than the radiation of the atmosphere. And I would put the two plots of the distributions of heavies to the different breakup models here. So we have 4 plots, all 4 about distribution of heavies near}}
	\label{DepletedMassPlanvsPebble}
\end{figure}

Next, we present the predicted heavy-element distribution in the planetary envelope as a function of time for different model assumptions.  
We define $f_{Z,env}$ as the fraction of the accreted heavy-element mass per time-step deposited at a given region in the envelope. 
In order to derive $f_{Z,env}$  we calculate the heavy-element mass accreted by the protoplanet $M_{Z,acc}$ at a given time-step and follow the ablation of planetesimals in the envelope. For simplicity, the calculated $f_{Z,env}$ corresponds for a given time, and is not affected by the distribution calculated at a previous time-step. 
We can then find the fraction of the accreted mass of solids (per time-step) that is deposited at different depths. %regions in the envelope. 
If all the planetesimals dissolve in the envelope, the integral of $f_{Z,env}$ over the planetary radius is equal to 1. At early stages when some planetesimals reach the core, the mass fraction of heavies that goes to the core is $1$ minus the integral. 
 %will be done in a future work where the evolution of the planet will be calculated self-consistently, keeping in account the presence of high-z material in the envelope. The convection will be studied with numerical analysis beyond mixing length theory} }%In other words, at a certain time-step $f_{Z,env}$ are not affected by the ones at previous stages, as the calculation is not self-consistent (this will be addressed in future works). 
In the left panel of Figure \ref{DepletedMassPlanvsPebble} we compare the inferred $f_{Z,env}$ assuming $C_h=0.1$ % with "instantaneous" fragmentation. 
 for planetesimals composed of rock+ice with sizes of $10$ m and $100$ km. 
$f_{Z,env}$ is shown vs.~normalized planetary radius.
In both cases at very early times, the heavy elements are deposited near the center (core), then the small planetesimals quickly stop reaching the core and their mass is deposited in the envelope. 
This also occurs for large planetesimals but with a time lag of $\sim $ 1 - 2 Myr, for our specific formation model.  
10 m-sized planetesimals enrich the outer part of the envelope and deposit most of their mass very far from the core, at normalized radius $0.7-0.8$. On the other hand, 100 km-sized planetesimals tend to enrich the inner parts of the envelope and most of their mass is deposited at a normalized radius of $\sim$ 0.2 - 0-3. In addition, $f_{Z,env}$ is found to be "smoother" for the smaller planetesimals. 
%{\bf Claudio, discuss the different panels... be clear and clean...explain what we see and what we learn from it...}\\
In both cases as time progresses, and the envelope mass increases, the heavy elements are deposited in the upper parts of the envelope (towards a normalized radius of one). 
%\st{The more "peaked distribution" of the large objects could change due to settling. are given in the Appendix. }
%Settling does not significantly change the distribution of the heavy elements. This is because settling is expected to be significant only in the outer part of the envelope where the temperature is below the critical value. 
%However, as can be seen in Figure \ref{DepletedMassPlanvsPebble}, planetesimal deposit the majority of their mass in the inner atmospheric layers (more details on settling are presented in the Appendix).

The sensitivity of $f_{Z,env}$ to the assumed $C_h$ value is shown in the middle panel of Figure \ref{DepletedMassPlanvsPebble}. %The left and right panels correspond to planetesimal sizes of 10 m and 1 km, respectively. 
We find that for the 10 m-sized planetesimal changing the $C_h$ value leads to a significant change in $f_{Z,env}$. A smaller $C_h$ value leads to a distribution with a peak closer to the core. 
When using a small $C_h$ value ($10^{-4}$), even for the small planetesimals there is a negligible enrichment in heavy elements in the outer parts of the envelope. % are found to be  also with small bodies the out part of the envelope is almost of solar composition
Since large planetesimals are less affected by ablation, the resulting $f_{Z,env}$ is less sensitive to the assumed $C_h$ value. 
In all the cases the heavy elements are deposited in the deep interior, leaving the outer envelope metal-poor. %depleted in heavy elements. 
%\textit{I have already talked about settling before, you think a repetition is good in this case?} \st{In addition, as discussed above, due to the more peaked distribution, the expect the heavy elements to sink to deeper regions..} \textbf{
The inner regions  are highly enriched mimicking a larger core. This configuration is consistent with a diluted core that is exact mass is not well-defined, due to the absence of a sharp boundary between it and the envelope (e.g. \citealt{Lozovsky17}, \citealt{Ravit17}, \citealt{Wahl2017}). It should be noted that for simplicity, we do not consider the re-distribution of heavies due to convective mixing (e.g. \citealt{Lozovsky17}, \citealt{Vazan2016}, \citealt{Venturini16}). 
If convection is efficient it would homogenize the gradient leading to a mixed envelope with a constant metallicity. 
Finally, in the right panel of Figure \ref{DepletedMassPlanvsPebble} we show the sensitivity of $f_{Z,env}$ to the assumed $\epsilon$ value. 
The trend is similar to the middle panel, but the dependence on $\epsilon$ is somewhat weaker implying that ablation is more important then radiation for these conditions. 
In the Appendix we present the results presented in Fig. \ref{DepletedMassPlanvsPebble} with $f_{Z,env}$ at different time in separate panels.
%This requires a self-consistent formation+evolution model, and we hope to address this in future work. 
%This is because in the inner regions is the frictional ablation that is dominant, due to the higher velocity of the planetesimal and density of the air.}

\subsection{Different Fragmentation Models}
\label{breakupsection}
%Here we focus on the location of the deposited mass, assuming so we see when the planetesimal is ablated, or broken, where it deposit its mass. Everywhere we will plot the ratio between the deposited mass and the total incoming mass, so the distribution for the deposited mass will be between $0$ to $1$. 

\begin{figure}[h!]
	\centering
	\subfloat[]
	{\includegraphics[width=.1532\textwidth]{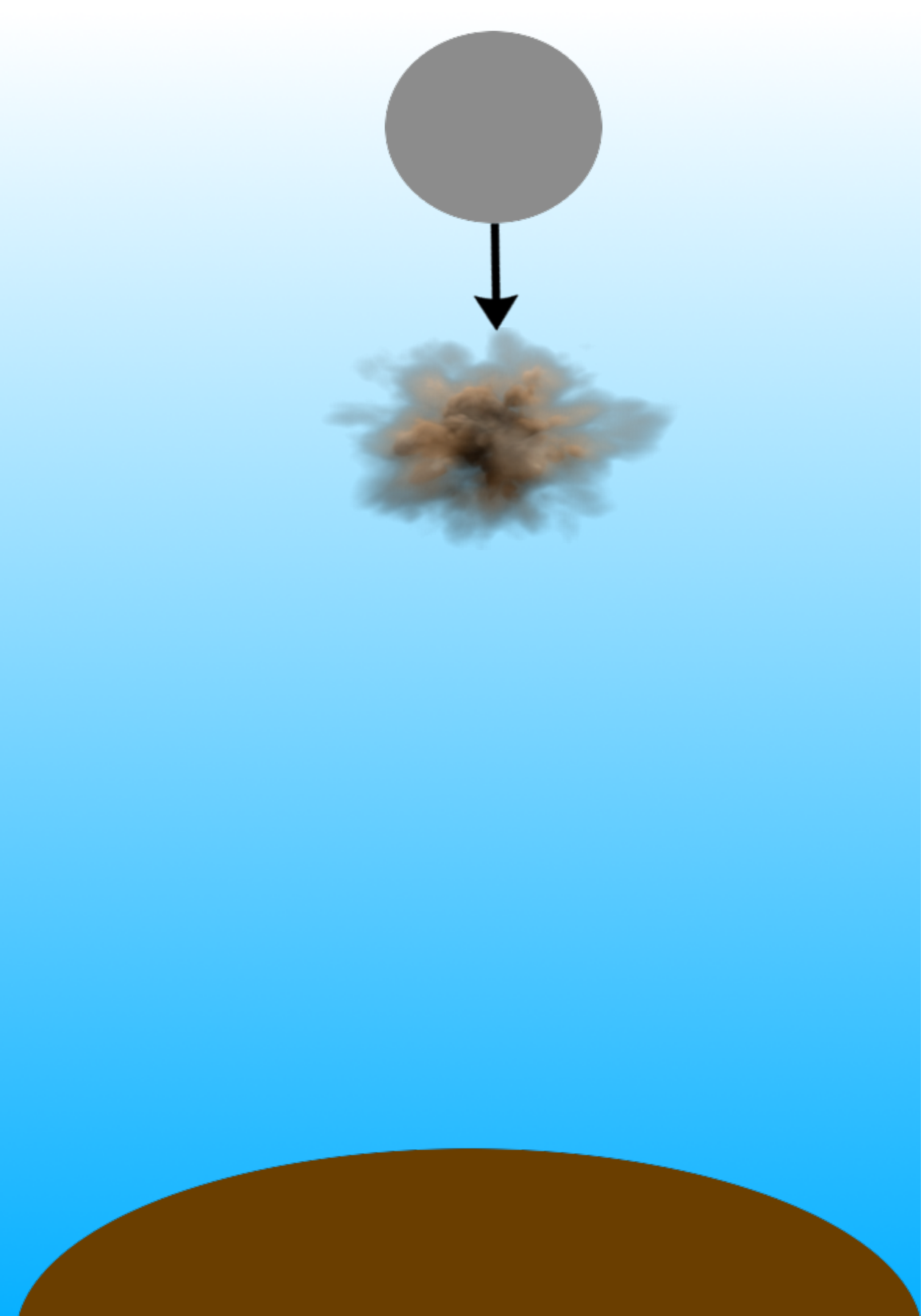}} \quad
	\subfloat[]
	{\includegraphics[width=.1532\textwidth]{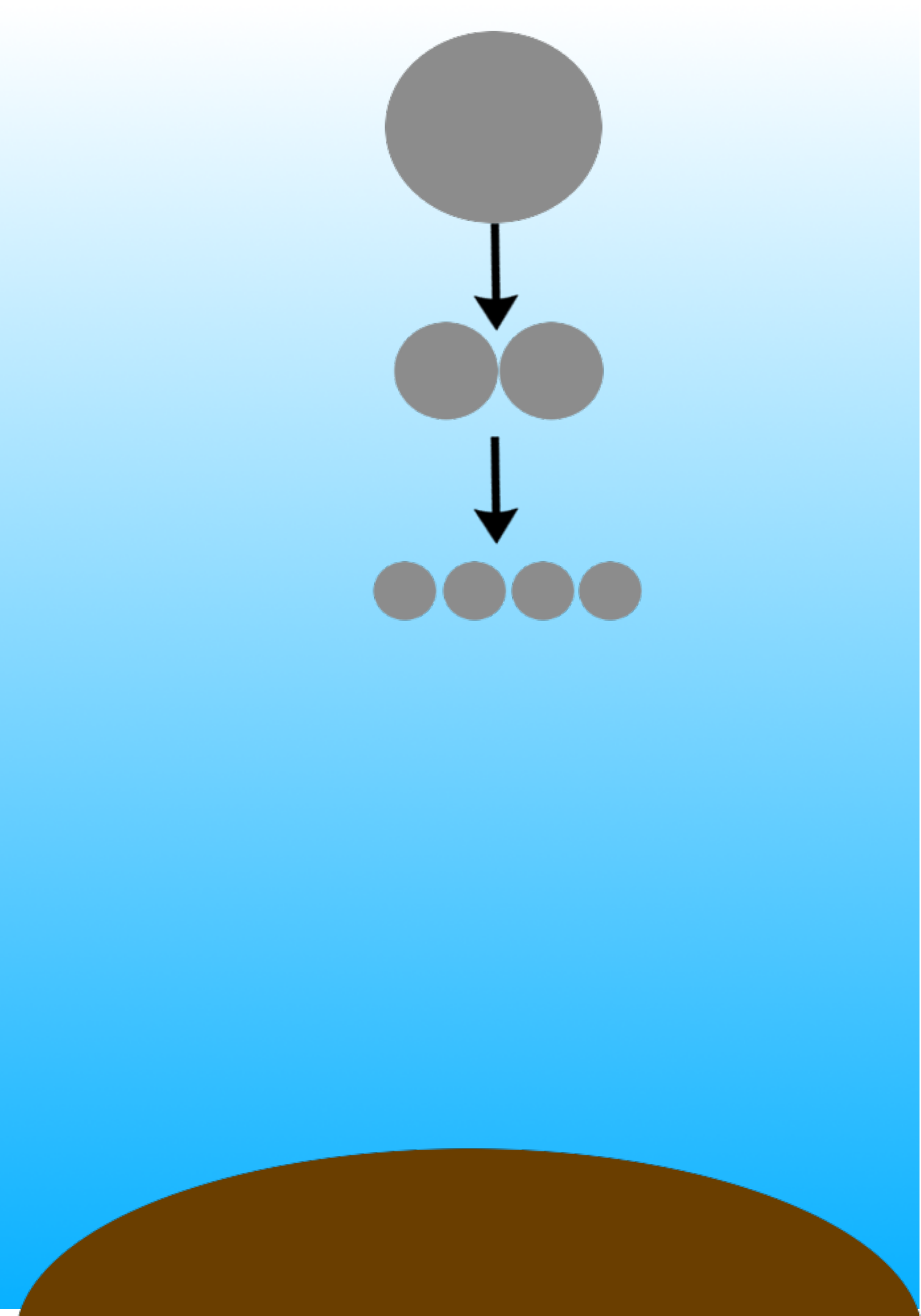}} \quad
	\subfloat[]
	{\includegraphics[width=.1532\textwidth]{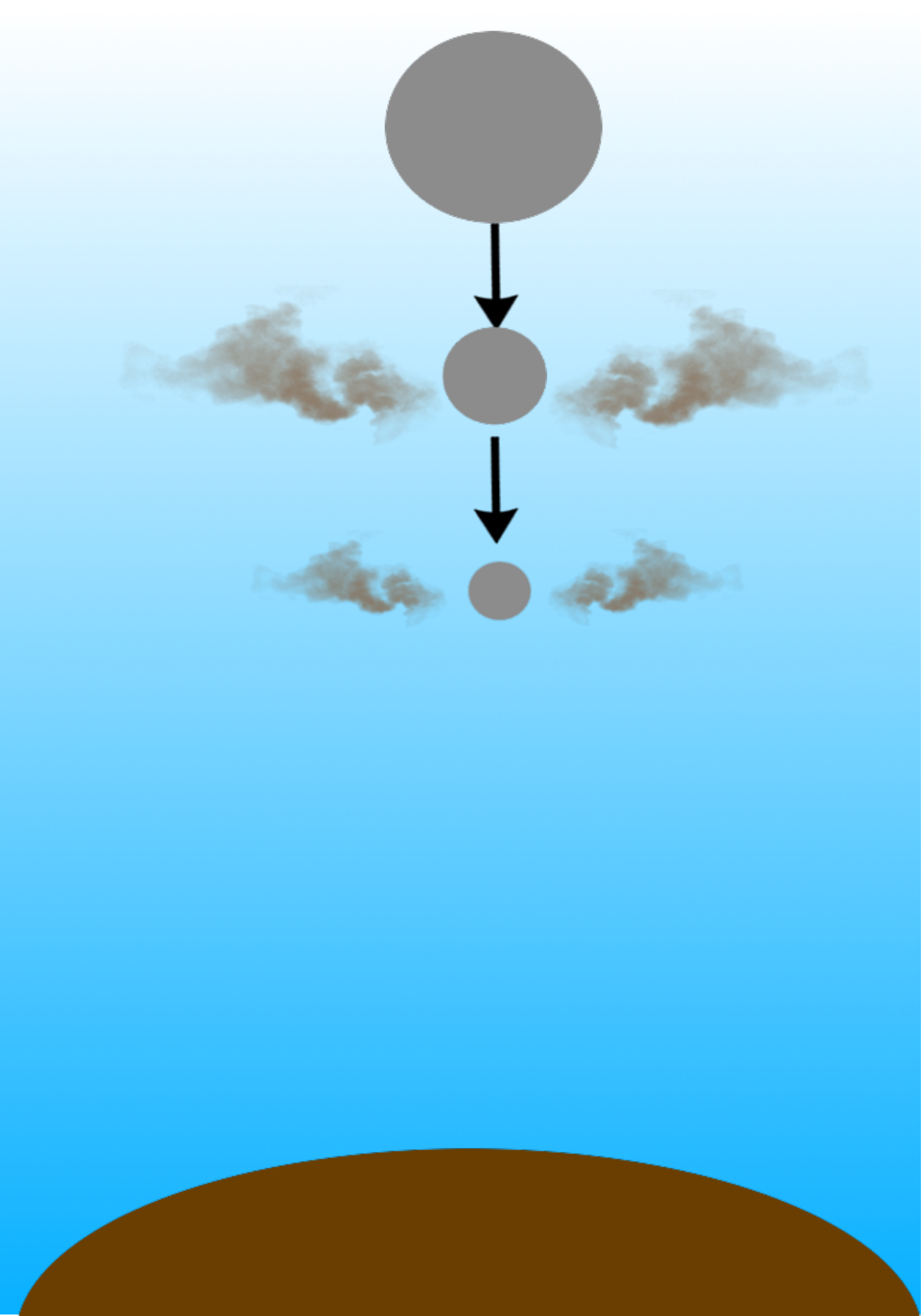}}  \quad
	\subfloat[]
	{\includegraphics[width=.1532\textwidth]{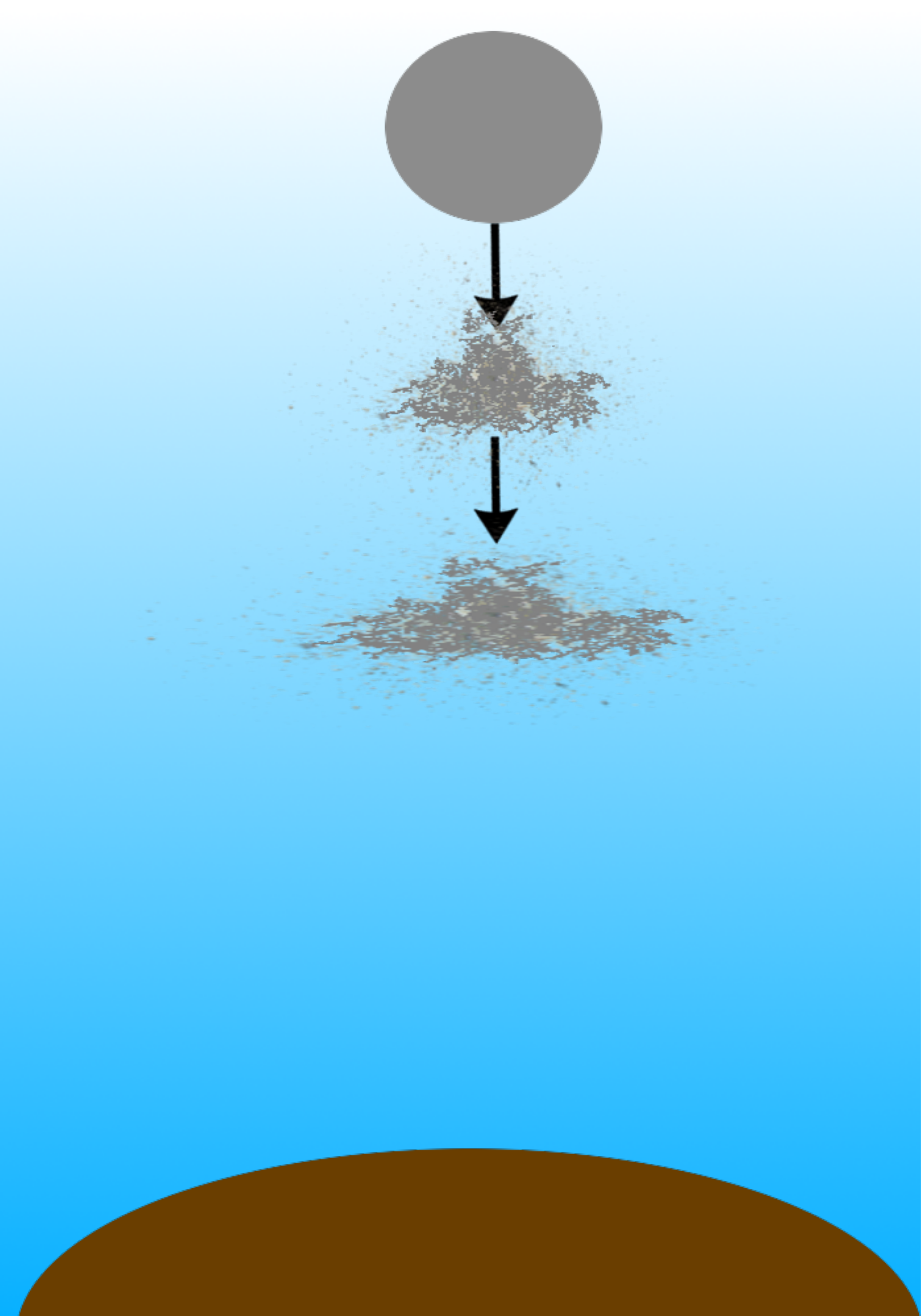}}  \quad
	\subfloat[]
	{\includegraphics[width=.1532\textwidth]{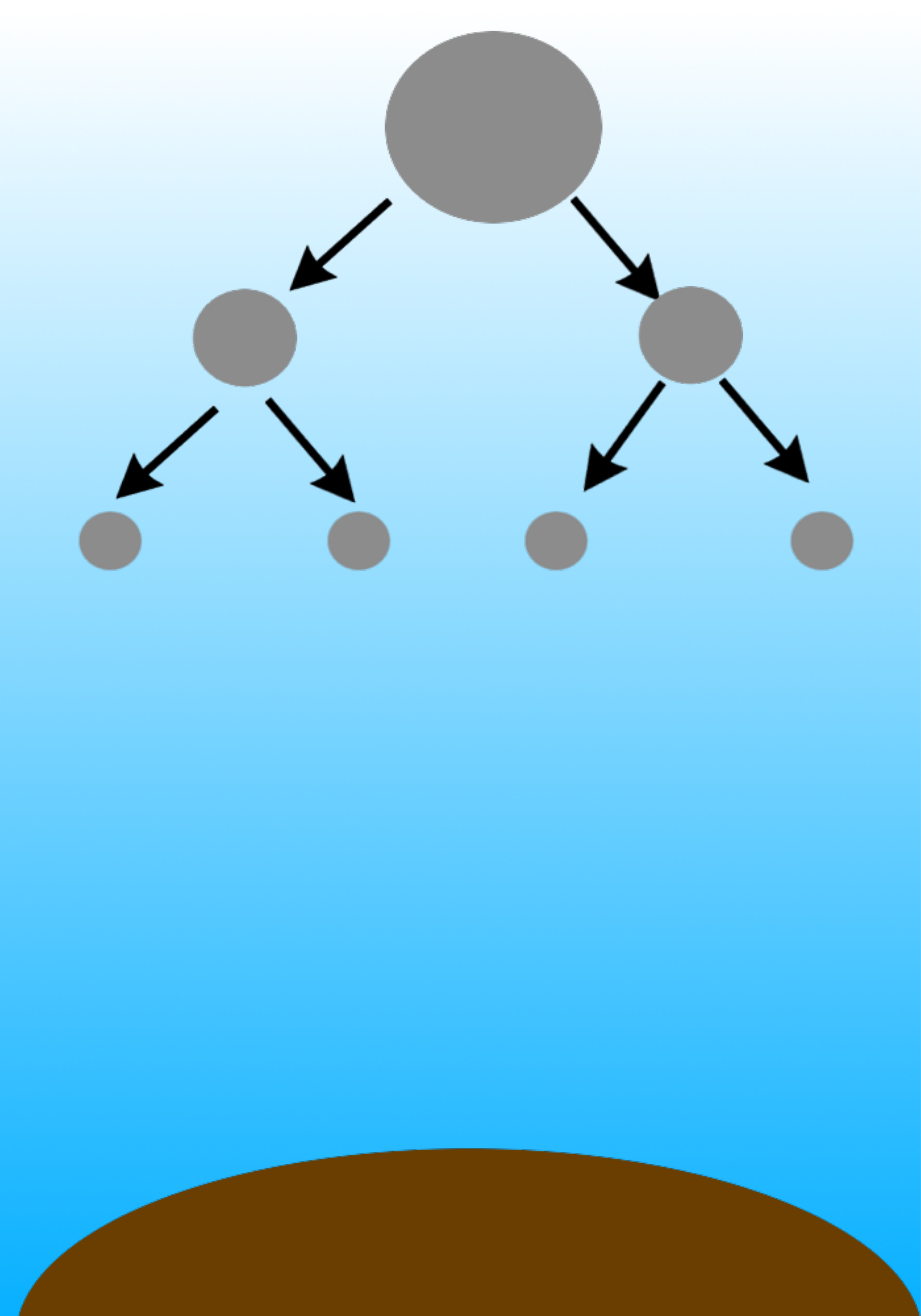}} 
	\caption{A graphic representation of the different fragmentation models: Instantaneous (a) Common Bowl (b), Halved (c), Pancake (d), and Single Bowl (e). More details on the fragmentation models can be found in \citep{Register2017}. }
	\label{breakupgraphic}
\end{figure}

When fragmentation occurs (Eq. \ref{breakupcondition}), it does not necessarily imply an instantaneous deposition of the entire planetesimal's mass at this location in the envelope. 
%The fragmentation event does not change the kinetic energy of the post-breakup material's mass \st{until the shape of it's bow shock has changed}. 
As discussed in \cite{Register2017} there are various ways to model fragmentation as listed below. A graphic representation of various fragmentation models is shown in Figure \ref{breakupgraphic}.

 % such as:  
%The simplest model is an instantaneous deposition of the material in the local layer (referred as "instantaneous"); more realistic models include: % model are summarized in \citep{Register2017} and are.
\begin{itemize}
     \sitem Instantaneous: Deposition of all the fragmented material at the layer where fragmentation occurs. This is the simplest model for fragmentation. 
	\sitem Single Bowl: The planetesimal fragments into two independent spherical bodies. Children bodies are assumed to be of equal mass,  half of the parent's mass and responds to gas drag independently  \citep{Mehta17}. The new size of the bodies is computed assuming that the same density as before fragmentation.  The bodies continue to move (following Eq. \ref{EqofMotion}) and lose mass due to ablation (Eq.~\ref{massablation}) with the surface term being adjusted accordingly.
	\sitem Common Bowl: When fragmentation occurs, the parent body is split into two equally sized fragments, each with half the parents' mass. The two fragments continue to interact with the gas being next to each other within a common bow shock, where the bodies present a common surface with respect to gas drag  \citep{Revelle07,Revelle05}.  This implies that  after fragmentation occurs the surface term in Eq.~\ref{EqofMotion} is the sum of the surfaces of the two children bodies.
	\sitem Pancake Model:  at the initial fragmentation point, the planetesimal is converted into a cloud of continuously fragmenting material that function aerodinamically as a single deforming body \citep{Zahnle92,Chyba93,Hills93}. 
	The cloud begins as a sphere and then flattens to get a pancake-like shape. The lateral spread is computed based on a dispersion velocity proportional to the square root of the envelope to planetesimal's density ratio and the instantaneous velocity: 
	\begin{equation}
		v_{disp}=({7 \rho_{gas}}/{2 \rho_p})^{1/2}v,
	\end{equation}
	 where $v$ is the impactor's instantaneous velocity.
	The area of the pancake is then calculated per each time step  $dt$ as	
	\begin{equation}
	A=\pi (r_{old}+v_{disp}dt)^2.
	\label{area}
	\end{equation}
	The material continues to move towards the proto-planet's centre following with the new surface $A$ given by Eq.~\ref{area}. 
	The drag coefficient calculated as in \cite{Podolak88} neglecting the non-spherical shape of the object \citep{Zahnle92}. 
	\sitem Halved: When fragmentation occurs half of the mass is assumed to be deposited at the local layer while the rest of the mass planetesimal continues to travel towards the center with the same velocity that it had before it fragmented \citep{Revelle07,Revelle05}.  
\end{itemize} 

 %of the planetesimal and we compare the result showing that when you compute the mass of the core, the different models using to describe the planetesimal break up, do not give very different result.

The material strength of the planetesimal is assumed to increase after fragmentation by the following power-law: 
$S_c=S_p(\frac{m_p}{m_c})^\alpha$ where $m_c$ and $m_p$ correspond to the mass of the child and parent, respectively, and $\alpha$ is a parameter between $0$ and $1$. 
Smaller fragments are assumed to have larger material strengths \citep{Mehta17,Artemieva01,Weibull1951}. The exact value of $\alpha$ is not well-determined because it depends on the inner structure of the planetesimal (both before and after fragmentation). In order to ensure that we do not bias the results, we run models with $\alpha$ values between $0$ and $1$ that are determined randomly, at each fragmentation.

\begin{figure}
\centering
\includegraphics[width=0.47\linewidth]{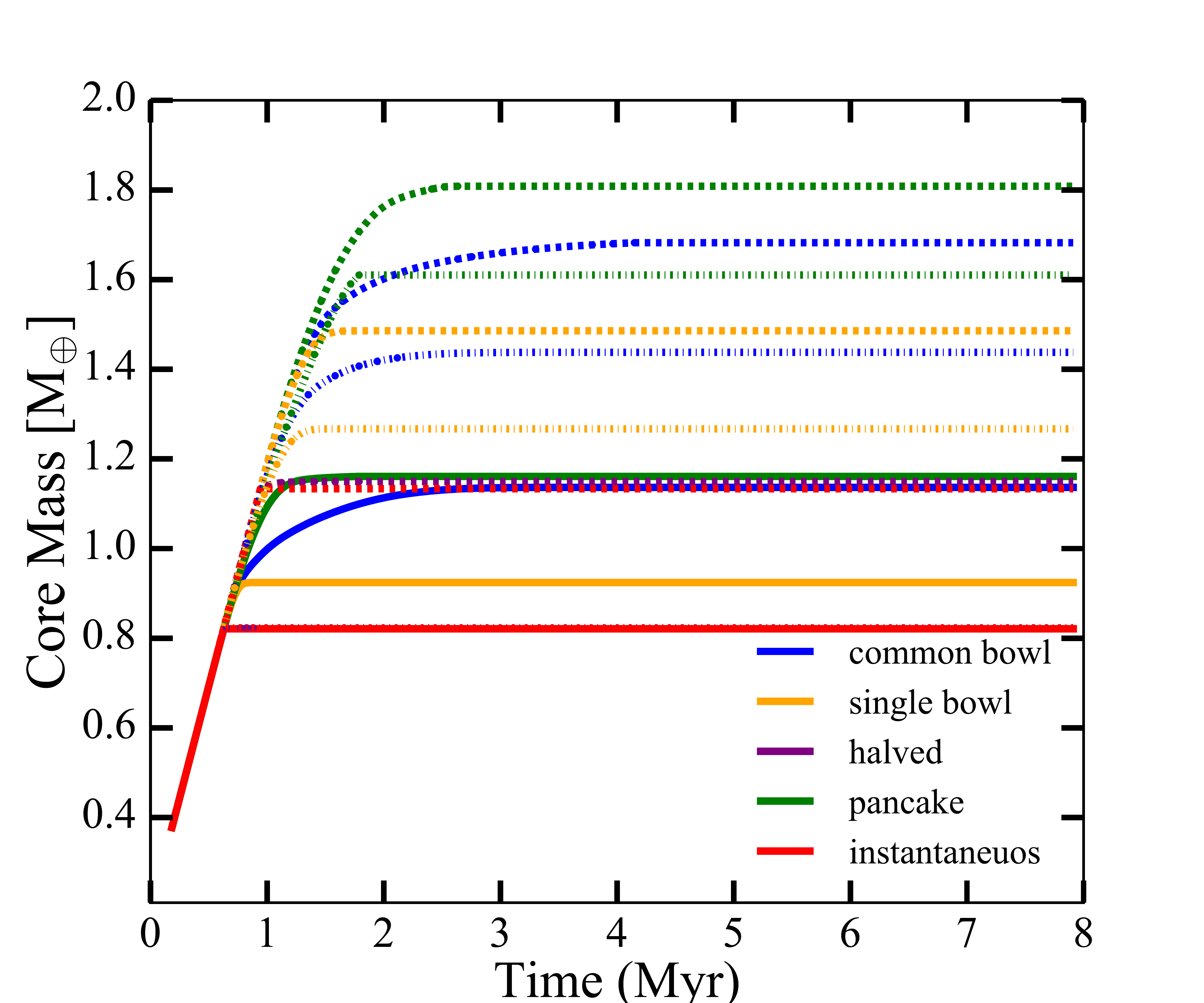}
\includegraphics[width=0.405\linewidth]{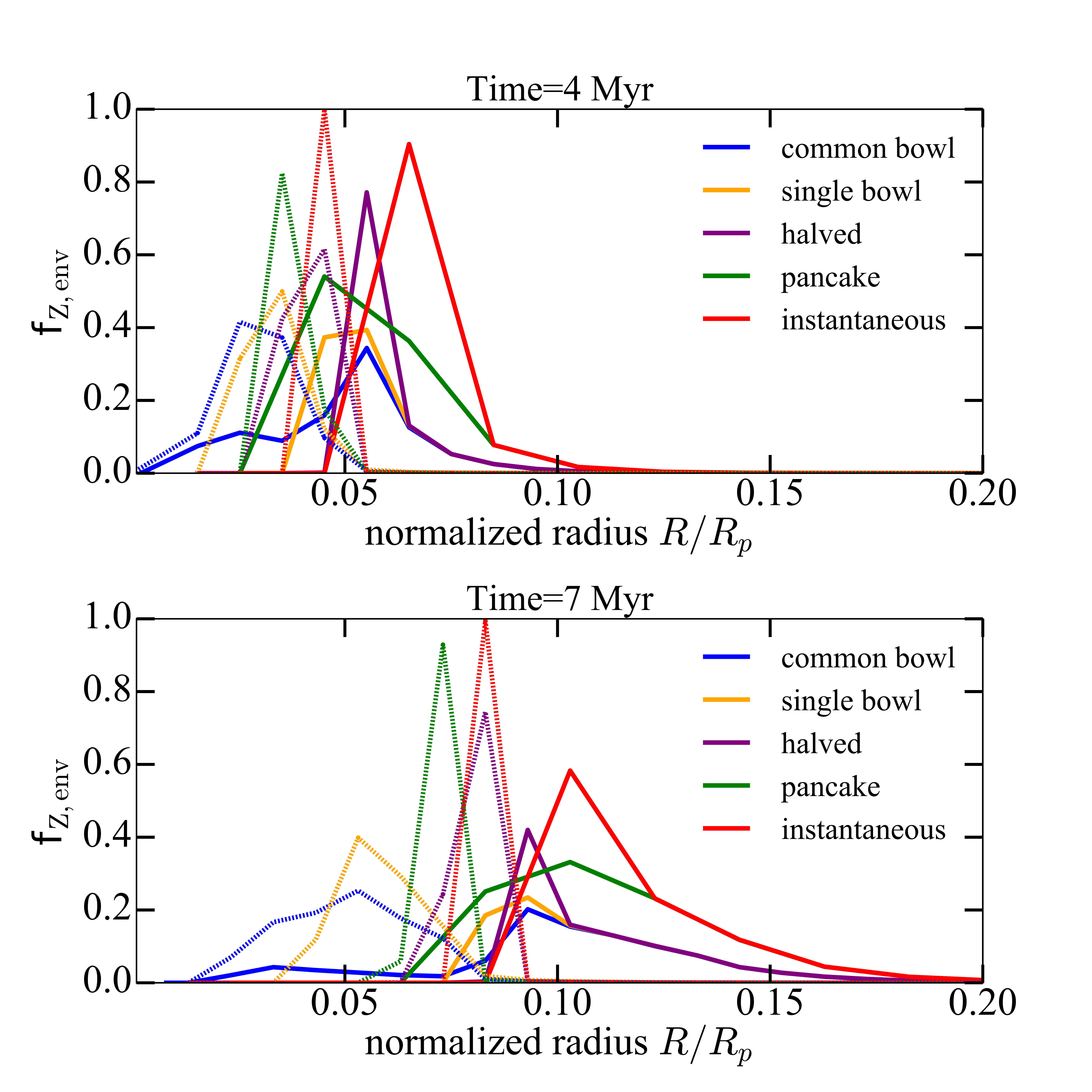}
\caption{{\bf Left:} Inferred core mass vs. time for the different fragmentation models. Solid and dashed-dotted lines are for 100 km-sized water planetesimals and $C_h =0.1$ and $C_h =10^{-3}$, respectively. % Dash-dotted lines are  is the same planetesimal with $C_h =10^{-3}$. 
The dashed lines correspond to 100 km-sized rocky planetesimals. {\bf Right:} $f_{Z,env}$  vs. normalized radius up to 0.2 at 4 Myr and 7 Myr corresponding to a total planetary mass of $5$ and $15$ M$_{\oplus}$, respectively.}
\label{Fig5}
\end{figure}

The left panel of Figure \ref{Fig5} shows the inferred core mass using different fragmentation models and $C_h$ values. Since small planetesimals typically do not fragment we consider only the cases of 1 km and 100 km. 
%\textbf{We find that the simple "instantaneous" predicts that core stops growing after 1 million year, reaching a core mass slightly below 1 M$_{\oplus}$. After this point planetesimals fragment near the core (due to the small envelope mass), and using different fragmentation model the core mass can increase by $0.1-0.5$ M$_{\oplus}$ within several thousand years.}
We find that the simple "instantaneous"  fragmentation model leads to the smallest core mass, smaller than 1 M$_{\oplus}$. 
Shortly after the point where the core stops growing in mass in the instantaneous fragmentation model, planetesimals fragment near the core (due to the small envelope mass). %, and allowing them to continue their journey keeping their kinetic energy, results in still hitting the core. 
When fragmentation is included, the core mass can increase by $0.1-0.5$ M$_{\oplus}$ within several thousand years. After that point even when fragmentation occurs the heavy elements are deposited too far from the core, and the core mass can no longer increase.   
\par

Changing $C_h$ can also slightly increase the core mass when the other fragmentation models are considered.  %but only for model different than the easiest instantaneous model. 
%\st{For example, the  green solid line, we can see that the dashed dot line (the line of the same planetesimal but lower $C_h$) is exactly over it. 
%This happens because $C_h$ does not alter the break-up condition, so when a planetesimal fragments, this is independent on the $C_h$ value. Changing the fragmentation model gives to the core mass a dependence on the $C_h$ value.} 
The "halved" fragmentation model predicts a core mass similar or slightly larger than the simplest "instantaneous" one, due to the rapid mass loss as more fragmentations occur. 
The largest core mass is obtained with the "pancake" fragmentation model due to the increased size of the "cloud" which travels towards the planetary center. Finally, similar core masses are predicted by "common bowl" and "single bow" fragmentation models. 
% with the former resulting always in an higher core mass. 
%In the "singe bowl" description, fragmentation leads to the formation of many small independent bodies that can be easily ablated by the gas, while in the "common bowl" scenario, "children planetesimal" move close to each other  in a compact formation that is hard to ablate.
%\par
%For example, we can analyze the pancake model. The solid line predicts a core mass of 1.1 M$_{\oplus}$ while the dashed-dot one is at 1.6 M$_{\oplus}$. Therefore, we conclude that the exact value of the predicted core mass depends on the assumed 
%fragmentation model, and the $C_h$ value. %Nevertheless, we find that the result that the core mass is of the order of up to 2 M$_{\oplus}$ is robust, and is independent on the fragmentation model. 
We conclude that the exact value of the predicted core mass weakly depends on the assumed fragmentation model, and is more affected by the assumed $C_h$ value. 
\par
%We will show the different output that these different model produces. We will show that when  the  fragmentation  takes  place  at  distances close  to  the  core, which is usually the case, the high-Z mass deposition is very localized so it is likely that any high-Z vapor in the impact plume will quickly become supersaturated and high-Z mass will sing to underlying layers. 
Next, we investigate the sensitivity of $f_{Z,env}$  to the assumed fragmentation model. Since small planetesimals are ablated, it is the large planetesimals that are affected by the treatment of fragmentation and we therefore concentrate on the distribution for planetesimals sizes of 100 km. The results are presented in the right panel of Figure \ref{Fig5} at two different formation times: 4 Myr (top) and 7 Myr (bottom). These times correspond to a total mass (core + envelope) of $5$ and $15$ M$_{\oplus}$, respectively. Note that we show the distribution up to a normalized radius of 0.2, since the distribution of heavy elements in outer regions is negligible since most of the material is deposited in the innermost regions.  
We find that $f_{Z,env}$ is relatively insensitive to the fragmentation model, but it does moderately affect the location of the peak and its spread. % for the different cases is somewhat different.  

\newpage
\section{A Semi-Analytical Approach to derive $f_{Z,env}$}
In this section we present a semi-analytical approach to derive $f_{Z_env}$ in the planetary envelope. 
The equation of motion can be solved analytically neglecting the contribution of gas drag, providing a simple semi-analytical solution for $f_{Z,env}$ . 
This can be applied to large planetesimals ($\geq$ 1 km) which are less affected by gas drag  (Equation \ref{EqofMotion}). 
%\st{With this approximation we can solve the equations analytically and provide a simple (semi-analytical) solution for the heavy-element mass distribution.}
%\par 
The equations can then be written as: %and we will plot result for the comparison
	\begin{equation}
 m_{pl}\frac{d\vec{v}}{dt}=-G\frac{M_{p}m_{pl}}{r^3}\vec{r},
\label{EqofMotion2}
\end{equation}
\begin{equation}
\frac{dm_{pl}}{dt}=-\frac{A}{Q}\bigg(\frac{1}{2}C_h\rho_{gas} \left| v\right|^3+\epsilon \sigma T_{a}^4\bigg),
\label{ablation2}
\end{equation}
where the gas drag term in Equation \ref{EqofMotion} is neglected. 
 The planetesimal's velocity is a 2D vector that can be decomposed as 
	\begin{equation}
		v=v_r \hat{r}+v_\theta \hat{\theta},
	\end{equation}
	where $v_r={dr}/{dt}$ is the radial velocity and $v_\theta$ is the angular velocity given by $r ({d\theta}/{dt})$, with $\theta$ being the polar angle. \\
Energy conservation implies:
\begin{equation}
v^2=v_0^2+2GM_{p}\bigg(\frac{1}{r}-\frac{1}{r_0}\bigg),
\label{velocity}
\end{equation}
where $v_0$ is the initial planetesimal's velocity and $r_0$ is its initial distance from the planet's center. Angular momentum conservation implies:
\begin{equation}
v_\theta^2=\frac{L^2}{m_{pl}^2r^2},
\end{equation}
where $L$ is the angular momentum, and the radial velocity is given by:
\begin{equation}
	v_r^2=v^2-v_\theta^2.
	\label{radialvelocity}
\end{equation}

Dividing both sides of Eq.\ref{ablation2} by $v_r$ results in: 
\begin{equation}
\frac{dm_{pl}}{dr}=-\frac{A}{Q}\bigg(C_h\frac{\rho_{gas} v^3}{2v_r(r)}+\epsilon \sigma T_{atm}^4\frac{1}{v_r(r)}\bigg), 
\label{dmdr}
\end{equation}
where $v$ is given by Eq.~\ref{velocity} and $v_r$ by Eq.~\ref{radialvelocity}.
Eq. \ref{dmdr} can also be written as: 
\begin{equation}
\frac{dm_{pl}}{dr}=-f(r), 
\end{equation}
where ${dm_{pl}}/{dr}=-\frac{A}{Q}\bigg(C_h\frac{\rho_{gas} v^3}{2v_r(r)}+\epsilon \sigma T_{atm}^4\frac{1}{v_r(r)}\bigg)$. 
%\end{equation}
Finally, the planetesimal's mass at a position $r$ within the envelope is given by 
\begin{equation}
\int_{M_0}^{M(r)}dm=\int_{r_0}^{r}f(r)dr,
\end{equation}
where $\rho$ and $T_{atm}$ are functions of $r$. The integration is performed up to M$(r)$ where $r$ is the location within the envelope where fragmentation occurs according to Equation (1). 
In order to estimate the total mass of heavy elements in the envelope one has to add the mass deposited due to fragmentation (the mass leftover after ablation). % $$

%We can now proceed in two way, or integrating this numerically, or try to intergate them analytically and obtain a full analytical approximation to the mass deposition of the planetesimal. If you want to follow the second approach you can write
%\begin{equation}
%\rho(x)=\rho_0e^{-x^2/H^2}
%\end{equation}
%\begin{equation}
%T(x)=T_0e^{-x^2/H^2}
%\end{equation}
%where $H$ is the scale height of the atmosphere (the average radius of the atmospheric layer). All the integrals that now compares can be done easily and you will get a rather lengthy expression. The only difficult integral to be handled is 
%\begin{equation}
%\frac{T_0e^{-4x^2/H^2}}{\sqrt{v_0^2+2GM_{planet}\bigg(\frac{1}{r}-\frac{1}{r_0}\bigg)}}
%\end{equation}
%Which I did not solved yet but has a solution.

\begin{figure}[h!]
  	\centering
  	\includegraphics[width=0.4251\textwidth]{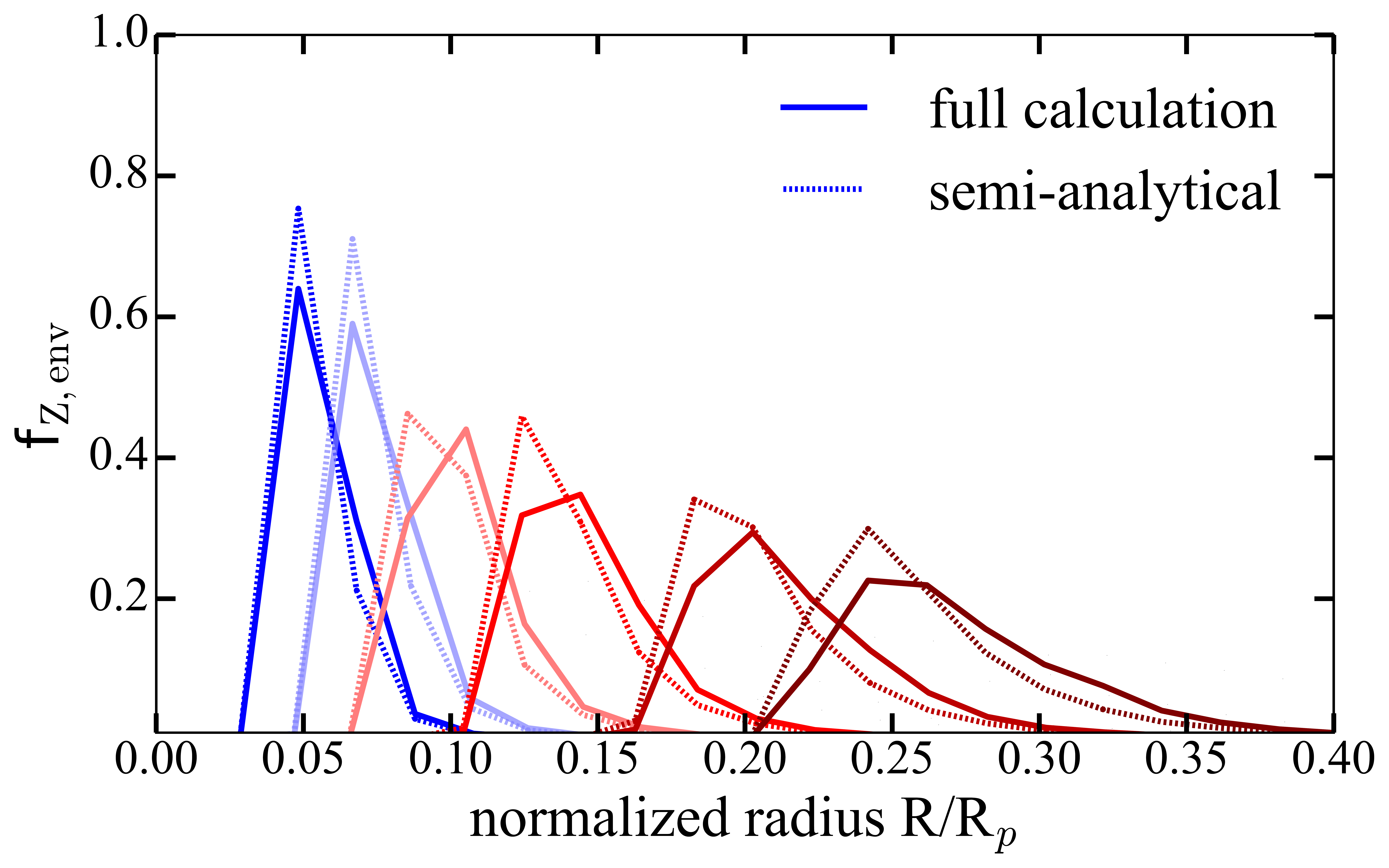}
	  	\includegraphics[width=0.4251\textwidth]{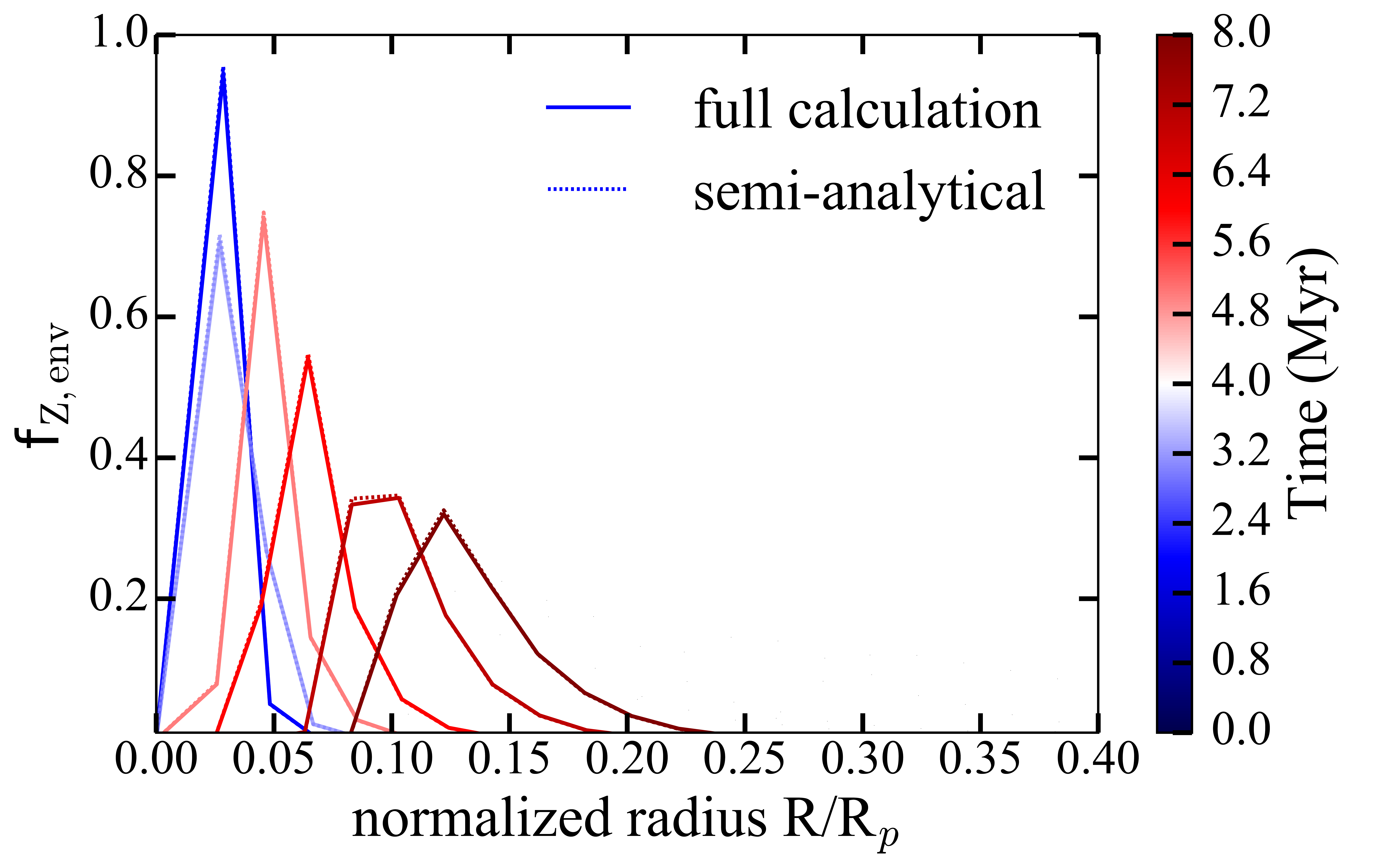}
  	\caption{A comparison between the numerical (solid curve) and semi-analytical (dashed curve) model for the inferred $f_{Z,{env}}$ at different times. Shown are the results for planetesimals made of rock+ice with sizes of 1 km (left) and 100 km (right).}
	% planetesimals made of water+rock. \textbf{Right:}  1 km-sized planetesimals made of water+rock. km size}
  	\label{fig:Time5}
  \end{figure}

We next integrate $f(r)$ numerically, starting from the core, simply using: % so the numerical integral was done with this formula
%\begin{equation}
$\int_{r_{core}}^{r_{max}}dr f(r)$ 
%\end{equation} 
where $dr$ is the size of an atmospheric shell and $f(r)$ is evaluated at the mid point in the atmospheric shell. 
%We use this semi-analytical formula to approxmiate t the ablated mass by the atmosphere, and the maximum mass that does not reach the core \ref{fig:Time4}. %An higher mass will hit the core.
%We next approximate the curve for mass depo s sition. 
The comparison between the numerical calculation and the analytical calculation is shown in Figure \ref{fig:Time5}. As can be seen from the figure, the agreement is excellent. % especially  In particular, as time progresses. 
We therefore suggest that this approximation can be used to generate the heavy-element distribution in porotoplanetary atmospheres in different planet formation models, including planet population synthesis models (\citealt{Mordasini2009}, \citealt{Benz14}).
 
%\newpage
\section{Discussion and Conclusions} \label{sec:Conclusions}
%{\bf Today, when the atmospheric composition of solar and extrasolar planets can be determined with spectroscopy measurements (see \cite{Mordasini16}), } and for connecting planet formation theories with internal structure models.
Predicting the heavy-element distribution in giant planets is crucial for the characterization of Jupiter and Saturn as well as giant exoplanets.  
Today, when the atmospheric composition of exoplanets can be determined with spectroscopy measurements, the determination of the expected heavy-element distribution in the planets' envelopes is vital to compare the measurements with theoretical predictions (e.g., \citealt{Helled14,Mordasini16}).
The internal structure of giant protoplanes depends on the interaction of the solids with the gaseous envelope during the early formation stages. 
%Constraining the distribution of the heavy elements in giant protoplanets is important and is not well determined. 
As we show here, the location in which the heavy elements are deposited depends on the planetesimals (solids) properties such as their size and composition, the fate of the accreted planetesimals (ablation/fragmentation) and other model assumptions. 
\par

We find that planetesimal fragmentation is important for planetesimal with sizes larger than 1 km, and that the importance of fragmentation vs.~ablation is sensitive to the assumed $C_h$ value. 
%The heavy-element distribution is found to be relatively insensitive to the applied fragmentation model. 
%The exact heavy-element distribution depends on the fate of the accreted solids (ablation/fragmentation). 
% whether planetesimal are ablated or broken by air pressure, 
Ablation typically results in a less-peaked distribution with enrichment of the outer regions of the envelope, while fragmentation enriches significantly the deep interior, leaving the outer envelope metal-poor. Finally, we present a semi-analytical prescription for determining the heavy-element distribution in the envelope. This can be easily implemented in giant planet formation calculation and can be used 
%This approximation is most appropriate for planetesimals with sizes of 1 km or larger where the contribution of gas drag to the equation of motion is negligible. 
%We suggest that future giant planet formation models should 
to include the presence of the heavy-elements in a more consistent manner, i.e., accounting for their effect on the EOS and opacity calculation.
\par

%The heavy-element distribution is found to be relatively insensitive to the applied fragmentation model, the assumed $C_h$ value and the expected radiation from the atmosphere, when ablation is dominating, can affect the results. 
% analyzing their impact on the final predicted core mass. We follow the distribution of heavies in the planet's envelope during its growth and we present the sensitivity of the distribution to $C_h$ parameter. Finally we present a semi-analytical formula to approximate the high-Z mass distribution in the proto-planet's envelope.
Although different model assumptions lead to different heavy-element distributions, we find that in all the cases heavy elements stop reaching the core once it reaches a mass of 0.5-2 M$_{\oplus}$. 
For our specific Jupiter formation model this corresponds to a time of 2 Myr. 
While the results presented in the paper such as the value of the maximum core mass, and the shape of the heavy-element distribution depend on the assumed atmospheric model and growth history, the general trend of the core reaching a maximum mass, and that most of the accreted heavy-elements 
are deposited in the envelope is robust and is in agreement with previous studies 
\citep{Lozovsky17,Brouwers2017,Venturini16,Mordasini2015,Iaroslav07}. 
\par
% are in agreement with ours, also if they use different atmospheric models.}
%This finding is robust and is %independent on the We find that this prediction is very strong, despite all possible changes of the model used the final core mass is very small, tipically between $0.5-2 M_{\oplus}$. This value is 
%in agreement with other studies using somewhat different approaches \citep{Lozovsky17,Podolak88,Brouwers2017}. 
It should be noted, however, that although the core mass is found to be very small, the inner region can still be highly enriched with heavy elements (nearly pure heavies). In that case the density profile is not very different from that of  a larger core, and this configuration can be viewed as a diluted/fuzzy core \citep{Lozovsky17,Ravit17,Wahl2017}. 
The core mass can slightly increase when fragments are allowed to reach the core but only by up to 0.5 M$_{\oplus}$. Also in this case the core mass does not exceed $\sim$2 M$_{\oplus}$. %but with a negligible value {\bf which one?}
We also show that dissolution of planetesimals in the envelope can increase the atmospheric mass (and its mean molecular weight) by a large factor.
\par

It is interesting to note that our study confirms the assumption of \citealt{Venturini16} that envelope enrichment begins once the core masses reaches a few Earth masses. While the exact number depends on the specific formation model and planetesimal properties, it supports the emerging picture that giant planets formed by core accretion are likely to have small cores and that envelope enrichment cannot be neglected. 
This work only explore the sensitivity of the inferred heavy-element distribution to different model assumptions. This is only the first step, and clearly more work is required. 
Future studies should also investigate the mixing of heavy elements at early stages to determine the expected structure of the envelope (homogenous vs. compositional gradients) and model the planetary growth in a self-consistent manner in which envelope enrichment is considered and is linked to the expected distribution of the heavy elements. 
%This is required since the final distribution of heavy elements depends on the mixing efficiency in the planetary envelope during the planetary formation and the long-term evolution. 
Accounting for the heavy-element distribution and their effect on the planetary growth and long-term evolution can improve our understanding of giant planet formation and  of the connection between the current-state structure and planetary origin. % and we hope to address this in tutu
%The expected distribution of heavy elements does not only affect the consequent growth but \textbf{is} also \textbf{important} for connecting the current-state with origin. 

\section*{Acknowledgments}
We thank Kevin Zahnle J., Morris Podolak, Julia Venturini, Yann Alibert and Allona Vazan for valuable discussions and suggestions. 
We also thank Christoph Mordasini for the careful and constructive reviewing of this manuscript.
%We also thank J. Venturini, Y. Alibert and C. Mordasini  for valuable discussions. 
R.~H.~acknowledges support from SNSF grant 200021\_169054. 
Part of this work  was conducted  within the framework of the National Centre for Competence in Research PlanetS, supported by the Swiss National Foundation.

\appendix
\section*{Planetesimal-Atmosphere Interaction}
\begin{figure}[h!]
	\centering
	\includegraphics[width=0.31582\linewidth]{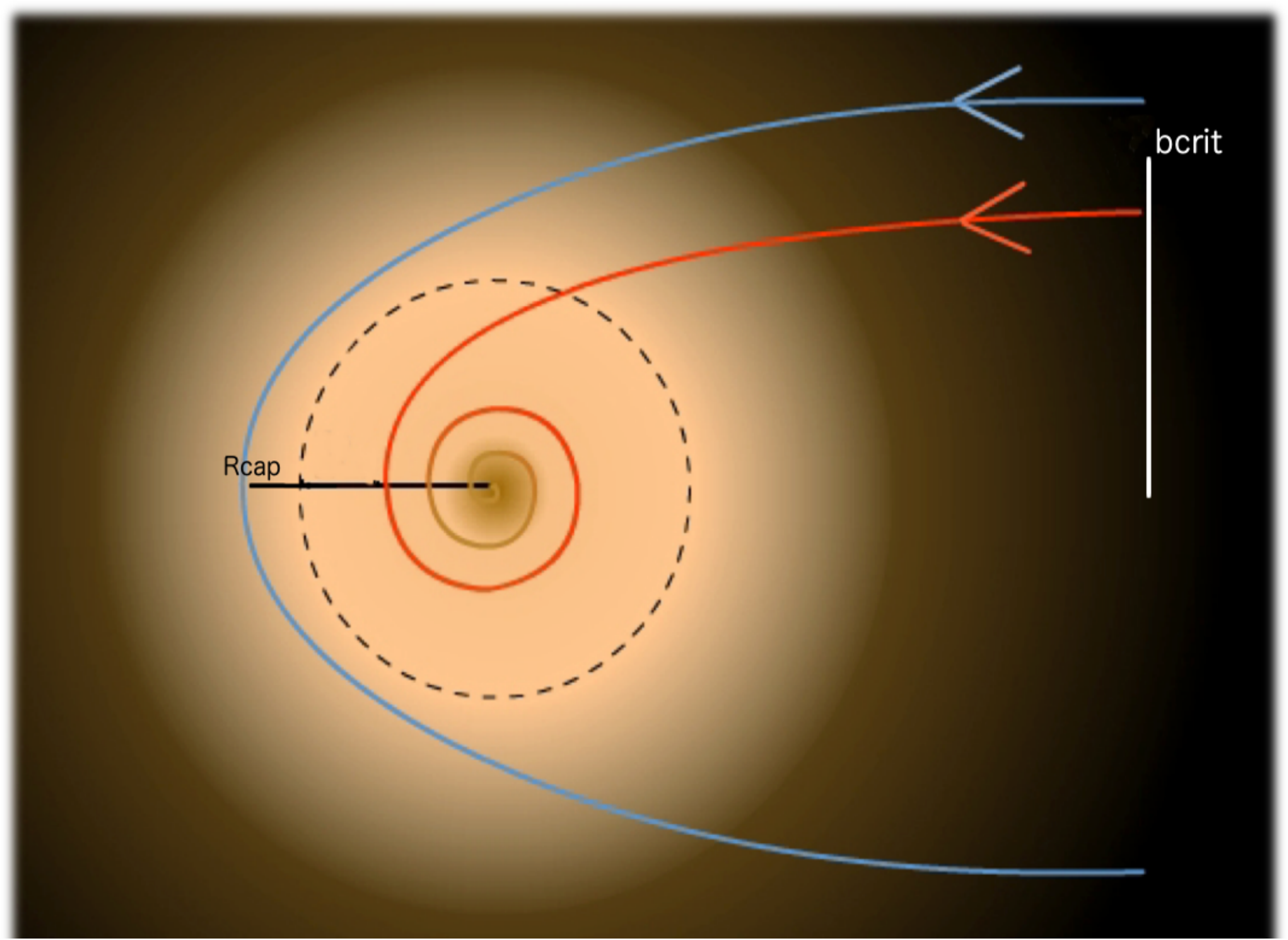}
	\caption{A graphic representation of the planetesimal-envelope interaction.}
%	
%		\st{The impact parameter of the incoming planetesimal is defined as the distance on the $y$ axis from the planet's center and the position of the incoming planetesimal.  $b_{crit}$ represents the maximum impact parameter that leads to capture. Any other planetesimal (with the same size and composition) with a higher $b$ escapes the planet. 
%$R_{cap}$ is defined as the closest distance to the planet's center for an impact parameter slightly great than $b_{crit}$.}}
	\label{GraphicRepresentation}
\end{figure}

%\subsubsection*{Basic definitions \& comparison with a  semi-analytical approach}
\label{appendix1}
The planetesimal dissolution is derived by following in detail the planetesimal's trajectory within the planetary envelope.  
%\st{interaction of the planetesimal with the gaseous envelope and  find the capture radius $R_{cap}$ of the protoplanet at a given time. In order to derive the capture radius (which changes with planetesimal size and composition) we follow the trajectory of a planetesimal within the planetary envelope.}
%\textit{You put the definition of capture radius and  bcrit in the caption of the figure, but I think is better to put in the main text rather than in the caption, what do you say? They are important things and is better to waste more space to explain their meaning, u agree?}
The planetesimal's impact parameter is defined as the distance on the $y$-axis between the initial point of the trajectory and the planet's center. 
The maximum impact parameter that leads to planetesimal capture is defined as the critical impact parameter $b_{crit}$.  
The value of $b_{crit}$ changes with the physical properties of the atmosphere (e.g., mass, pressure, temperature) and the planetesimal (e.g., size, composition).
% physical properties. depends on the charateristic of the giant planet and the planetesimal nature.  
A planetesimal (with the same size and composition) with an impact parameter larger than $b_{crit}$ escapes the planet and cannot be accreted. 
 Planetesimal escape is defined to occur when the planetesimal's kinetic energy is slightly greater than the gravitational one as defined by \cite{P96}. % by the planet. 
The planetary capture radius $R_{cap}$ is defined as the closest distance to the planet's center for an impact parameter slightly great than $b_{crit}$ \cite{Ravit06}. 
Figure \ref{GraphicRepresentation} presents a graphic representation of $b_{crit}$ and $R_{cap}$. 
%{\bf Typically, the planetesimals are assumed to be uniformly distributed among the different impact parameters (i.e., between $[0,b_{crit}]$), and the work presented above adapted this assumption. 
%However, in order to make sure that the distribution of heavy elements within the protoplanet does not depend on the distribution of planetesimals within the range of impact parameters that lead to capture we also consider a case in which the planetesimals have a  gaussian distribution picked at the mid-plane. }
%\textbf{An uniform solid population of the region $[0,b_{crit}]$ was assumed in all the result presented in this work . In \ref{GaussianAppendix} are shown results for a gaussian population distribution.
%It is clear that if planetesimal ablation is neglected $b_{crit}$ depends only on the mass of the planet. When considering gas drag and ablation of solid $b_{crit}$ becomes dependent on the nature of the planetesimal, the density and mass of the atmosphere. Considering ablation affects significantly the envelope mass. \ref{AtmoMass} shows the difference of the core mass and the envelope mass with and without ablation}
In Figure 10 we compare our inferred $R_{cap}$ with the semi-analytical formula of \cite{Inaba03}. 
%The comparison for four different atmospheric models is shown in figure \ref{JapaneseFig}. 
As can be seen from the figure, there is a very good agreement. 
\begin{figure}[h!]
	\label{JapaneseFig}
        \centering
        \includegraphics[width=0.9\linewidth]{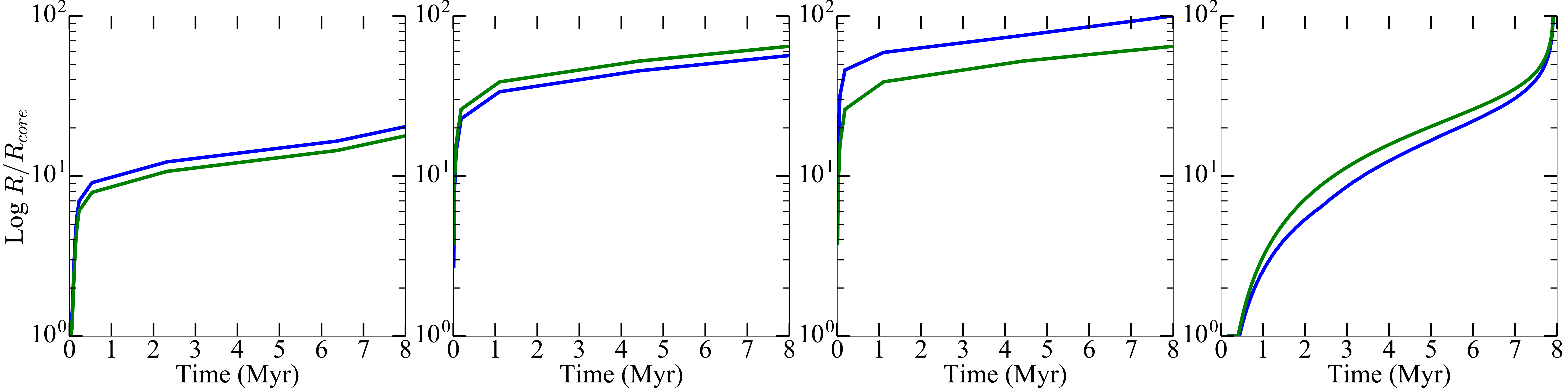}
        \caption{ A comparison of the inferred $R_{cap}$ using our numerical calculation (blue line) with the semi-analytical formula of \cite{Inaba03} (green line) for four different formation models. 
        Left, middle left and middle right panels correspond to models with a surface solid density of $10$ $g/cm^2$ and planetesimal sizes of $100$ km, $1$ km, and $10$ m, respectively. 
        Right: the formation model used in this work with $100$ km-sized planetesimals.}
        \label{japanese}
\end{figure}

 \newpage
\section*{The Distribution of Impact Parameters}
\label{GaussianAppendix}

\begin{figure}[h!]
	\centering
	\includegraphics[width=0.35\linewidth]{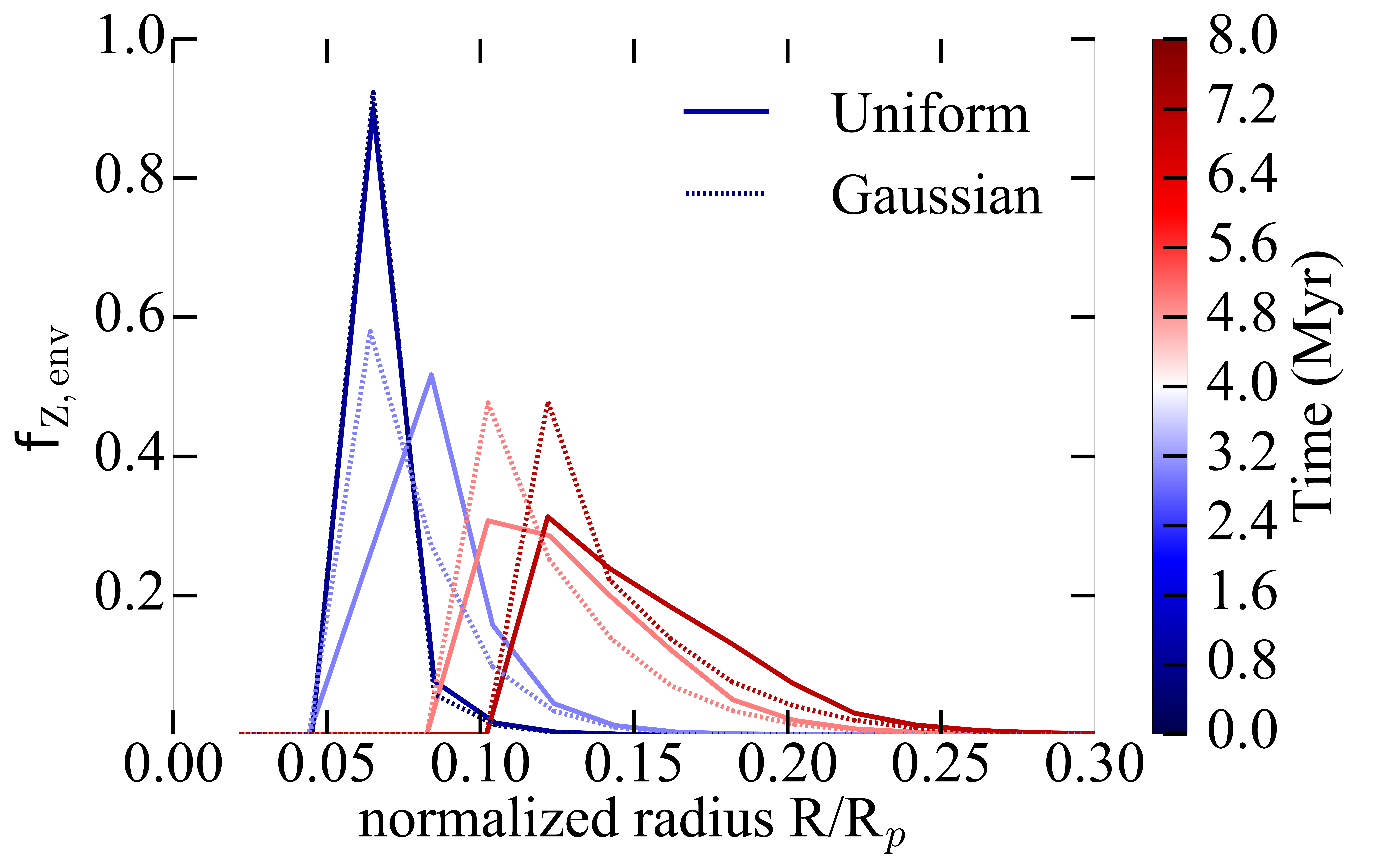}
	\caption{$f_{Z,env}$ at different times using two different distribution of planetesimals along the planet's feeding zone. The solid line correspond to a uniform distribution while the dotted line for a gaussian distribution picked at the mid-plane. The planetesimals are assumed to be pure rocky and 100 km in size.}
	\label{Gaussian}
\end{figure}

In order to infer the heavy-element distribution we distribute $100$ planetesimals with impact parameters between zero and $b_{crit}$. 
Planetesimals with large impact parameters tend to spend more time in the outer part of the envelope and to experience gas drag and ablation before they are captured. 
%, and they will feel more the ablation and gas drag. {\bf Claudio, please explain?!}
%Their distribution will be more uniform and more focused on the outer layers on the envelope. 
On the other hand, planetesimals with small impact parameters can penetrate deeper in the atmosphere and reach closer to the core. 
We investigate the sensitivity of the inferred heavy-element distribution in the planetary atmosphere on the assumed impact parameter distribution. 
We consider two cases. In the first case we assume that the planetesimals are uniformly distributed along the z-axis (uniform) while in the second one we assume a gaussian distribution for the impact parameters picking at the disk's mid-plane (planetary equator). 
%It is reasonable in fact, that the distribution over the impact parameter is not uniform but actually follows a Gaussian distribution, with a peak in the centre of the disk. 
The width of the gaussian distribution is set to be time dependent ranging from 0.5 to 0.1 of the disk's height, which is taken to be the ratio between the sound speed and the Keplerian velocity.
Such a distribution is more realistic since the concentration of planetesimals is expected to increase toward the disk's midplane. 
% this to represent the settling of the planetesimal towards the mid plane of the disk.
A comparison of the inferred $f_{Z,env}$ assuming uniform and  gaussian planetesimal distribution is presented in Figure \ref{Gaussian}. %population of the region $[0,b_{crit}]$ and a Gaussian one is shown in Figure \ref{Fig15-Gaussian}. 
We find that the gaussian distribution results in a slightly higher concentration of planetesimals in  the deep interior. This is an expected result because in that case there is a larger number of planetesimals at the disk's mid-plane that can reach the central regions. % and they can get closer to the planet's center leading to a higher peak in the heavy-element mass fraction  near the center.
However, despite this difference, the two curves are similar and differ by only up to 5-10\%. We therefore conclude that for the purpose of calculating the heavy-element distribution in protoplanets assuming a uniform distribution is sufficient.  
% In addition, we find that the gaussian distribution results in a slightly higher concentration of planetesimals in  the inner regions. 
%%                                                                    rather than an uniform one is that the peak is a bit higher. 
%This is because there are more planetesimal near the planet's equator and they can get closer to the planet's center leading to a higher peak in the heavy-element mass fraction  near the center.

%\section{Results}
\section*{The Importance of Settling}
\label{Settlingsection}
When heavy elements dissolve in the envelope they are assumed to be fully vaporized. 
Then, the  amount  of  heavy elements  that  remains  at  a  given  location within the envelope is determined by the local temperature and its vapor pressure $P_{vap}$. 
In order to calculate how much of the ablated mass remains in the given shell we  follow the procedure presented in \cite{Iaroslav07} 
and assume that the entire ablated mass is initially in the vapor phase, and then compare the partial pressure $P_{par}$ of the vapor to the saturation vapor pressure at the ambient temperature. If  $P_{par} >  P_{vap}$, the abundance of the volatile mass in the gas phase is such that $P_{par} =  P_{vap}$, and any excess of material settles to the layer below. 

In computing the vapor pressure we assume that the vapor obeys the ideal gas law. % (see Iaroslavits \& Podolack 2006 for details).
The expressions for the vapor pressure are  
$P_{vap}^{ice}=e^{-5640.34/T+28.867}$ and $P_{vap}^{rock}=10^{-24605/T+13.176}$ (see \citet{Iaroslav07,Lozovsky17} for details). %taken from  \textit{CRC Handbook of CHemistry and Physics}. 
%Modelling accurately the heavy-elements at the critical point is very important. 
In this approach, it is assumed that unlimited amount of heavy-element material can be kept in the vapor phase in regions where the temperature exceeds the critical point. 
The critical temperature can be derived from experimental data and are taken to be 567.3 K for H$_2$O and 4000 K for rock.	 
\par

\begin{figure}[h]
	\label{Settling}
	\centering
			\includegraphics[width=0.32\linewidth]{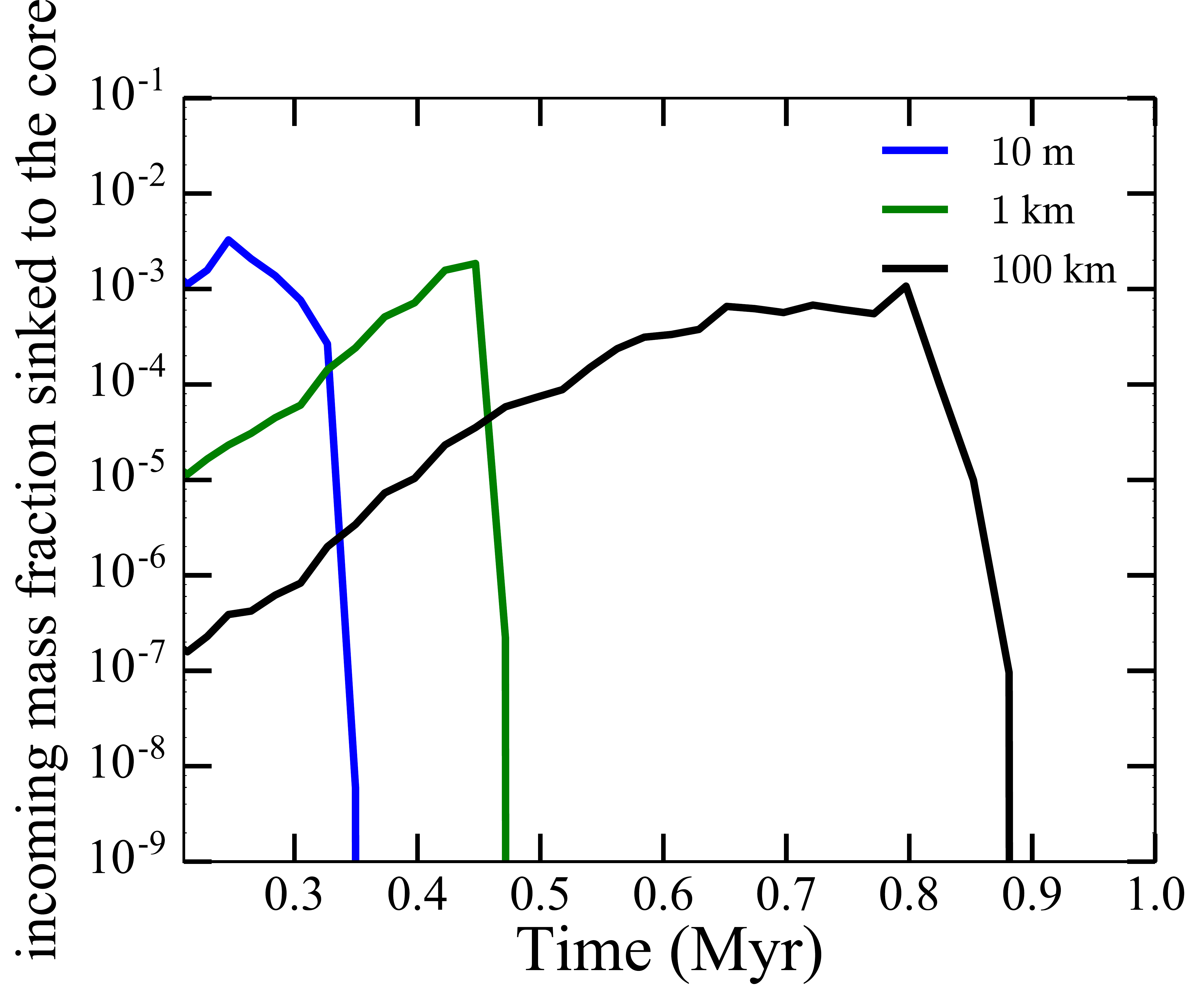}
	\caption{
	%Left: Partial pressure over vapor pressure for a water planetesimal, early times (1 My) and late times (7 Myr). Right: 
	Mass added to the core by settling for planetesimals made of water+rock. We consider three different planetesimal sizes: 10 m (blue), 1 km (green) and 100 km (black). 
	%Here the planetesimal a planetesimal of water+rock (50-50 composition) - what radius {\bf ??? Caludio?}}
	}
	\label{SettlingCoreMass}
\end{figure}

At each timestep we begin from the outermost layer and follow the material that settles to the core. 
Figure \ref{SettlingCoreMass} shows the mass that is expected to be added to the core via settling. 
%We plot also the total solid mass that is being accreted by the planet. 
We find that the mass joining the core due to settling  is very limited and can be neglected. 
This result, however, is linked to the assumption that hot enough (above the critical temperature) regions can absorb the heavy-elements. Future studies should include the formation of clouds and follow the settling of elements at the critical point in order to provide more robust conclusions on the importance of settling inside giant protoplanets. 
% correspond to the way we treat the vapor in regions where the temperature exceeds the critical point. 
%In this approach an unlimited amount of heavy elements can be absorbed as vapor phase for temperature above the critical point result 
%in turning off settling in the inner layer, where the deposition of heavies by the planetesimal is more significant. In addition to that,} 
%the dependence on the temperature of the vapor pressure is exponential and therefore,  the amount of material that can be "absorbed" by the envelope in the gaseous phase is increasing with depth. 
%%, preventing material to fal the core. These result are in agreement with  \cite{Lozovsky17} 
%%thanks to sinking of dust in layer is up to two order of magnitude less than the total solid mass that is being accreted by the planet. 
%As a result, we conclude that the increase of the  core mass by settling is very limited and can be neglected. 
%
%In Figure \ref{Settling} shows the ratio of the partial pressure over vapor pressure. When this quantity is greater than $1$ sinking takes place. When it is below $1$ all the ablated mass can be mantained in the gaseous phase. We can see that the ratio is greater than one only in the outern part of the envelope, while planetesimal deposit the majority of their mass in the inner layers. 
%We can thus neglect settling under this assumption and consider, as the greatest source of indirect core growth the different breakup models. 

%\begin{figure}
%	\centering
%	\includegraphics[width=0.5\linewidth]{Fig17-Pressure}
%	\caption{Partial pressure over vapor pressure for a water planetesimal, early times ($1 \times 10^6$ years) and late times $7 \times10^6$ years}
%	\label{VaporPressure}
%\end{figure}

%\clearpage
\section*{A different representation of $f_{Z,env}$}
\label{DifferentF}
%{\bf Here we show.... move what is written above in the text down here but then mention this appendix - also, check if ApJ appex should have numbers and then fix the text so it is consistent.... 
%don't forget to check all the variables in the paper!!! and look at the figures - they should look good!!!, and change to $f_{Z,env}$ instead of distribution of heavy elements"... etc.
%and change the colors and style of the figures below so they are consistent with the rest...}
%
 Below we the results of Fig.~5 for $f_{Z,env}$ when the various times are presented in separate panels. The left panel of figure 13 shows  $f_{Z,env}$ when assuming rock + water planetesimals with different sizes. 
It can be seen that larger planetesimals tend to enrich the inner part of the envelope, while small planetesimals deposit their mass in the outer regions. 
In the right panel of figure  we present $f_{Z,env}$ for different assumed $C_h$ values. As expected, a smaller $C_h$ value leads to a distribution with the peak being closer to the center.

\begin{figure}[h!]
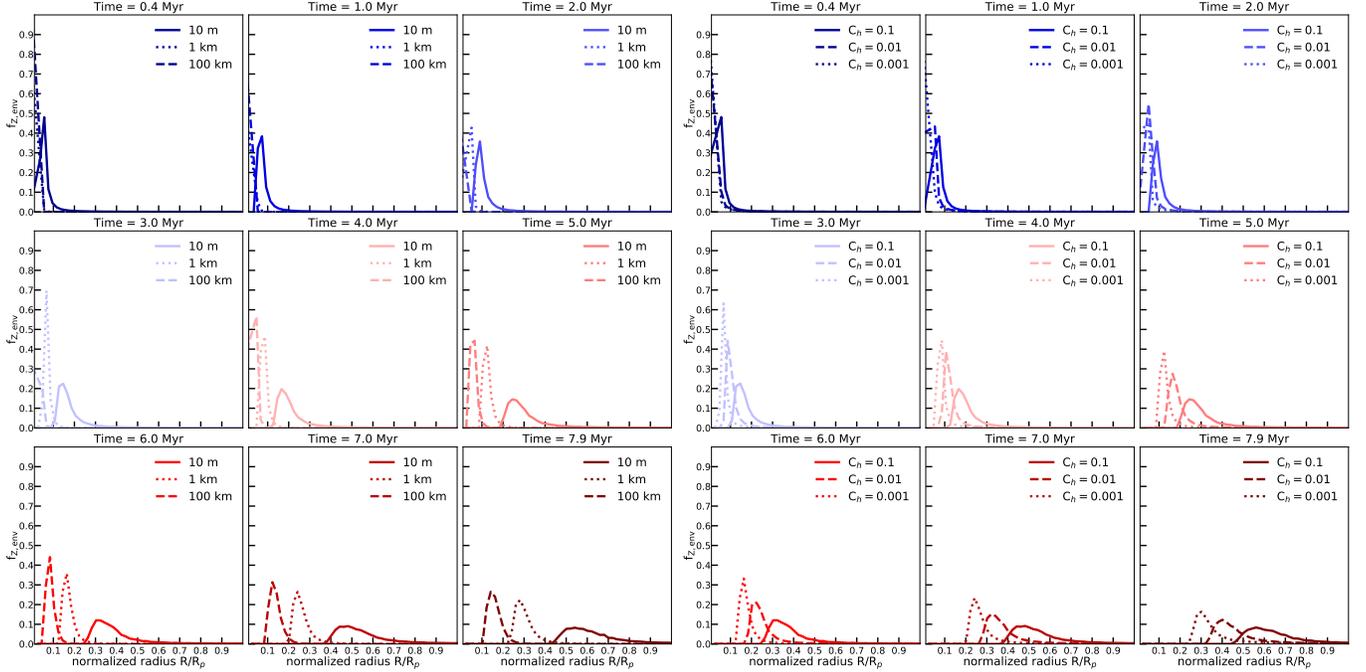

	\centering
	\includegraphics[width=0.49310\linewidth]{Depleted_Mass_2}
		\includegraphics[width=0.49310\linewidth]{Depleted_Mass_9Plots}
	%\vskip{-0.2cm}
	%\vskip -25pt
	\caption{$f_{Z,env}$ in the planetary envelope vs. normalized radius at different times. The times are shown in different panels. Left: Results for different planetesimal sizes. 
	The solid, dotted and dashed lines correspond for planetesimals with sizes of $10$ m, $1$ km and $100$ km, respectively. 
	Right: Results for different assumed  $C_h$ values. The solid, dashed and dotted lines correspond for planetesimals with a $C_h$ value of $0.01$ and $0.001$, respectively. 
	The planetesimals are assumed to be made of rock and be $10$ m in size.}
	\label{BigFigurech}
\end{figure}

%%\newpage
%\begin{figure}[h!]
%	\centering
%	\includegraphics[width=0.550\linewidth]{Depleted_Mass_9Plots}
%	%\vskip{-0.2cm}
%	%\vskip -25pt
%	\caption{{\bf $f_{Z,env}$ in the planetary envelope vs. normalized radius at different times. The times are shown in different panels. Planetesimals are assumed to be rocky with sizes of $10$ m.
%	The solid, dashed and dotted lines correspond for planetesimals with a $C_h$ value of $0.1$, $0.01$ and $0.001$, respectively.}}
%	\
%{BigFigDifferentSize}
%\end{figure}


\clearpage

\begin{thebibliography}{}
	
	\bibitem[\protect\citeauthoryear{{Alibert}}{{Alibert}
	}{2017}]{Alibert17}
	{Alibert}, Y.,\ 2017, \aap, 606, A69
	
\bibitem[\protect\citeauthoryear{{Alibert} et~al.}{{Alibert}
  et~al.}{2005}]{Alibert05}
{Alibert}, Y., {Mordasini}, C., {Benz}, W.,  \& {Winisdoerffer}, C. 2005, \aap,
  434, 343
  
    %Cited
    \bibitem[\protect\citeauthoryear{{ALLEN, H. J.}}{{Allen, H. J. }}{1962}]{Allen1962}
    {{Allen, H. J.}}, 1962, NASA SP-24.
    %Cited

        %Cited
        \bibitem[\protect\citeauthoryear{{Apstein E.Z.},{Pilyugin N.N.P., S.},{Evastianenko V.G.},{G.A. Tirsky}}{{Apstein et~al.}}{1989}]{Apstein1989}
        {{Apstein, E. Z.}, {Pilyugin N. N. P. S.},{ Evastianenko V.G.},{ G.A. Tirsky}} 1989, Itogi Nauki Tekh. Mekh. Zhidkosti Gasa, 23, 116–236.    
        %Cited
        
          %Cited
          \bibitem[\protect\citeauthoryear{{Artemieva, N.A.},{Shuvalov, V.V.} }{Artemieva et~al.}{2001}]{Artemieva01}
          {Artemieva, N.A.},{ Shuvalov, V.V.} 1996, Shock Waves, 5(6), 359-367
          %Cited
          
                    %Cited
          \bibitem[\protect\citeauthoryear{{Baraffe, I.},{Alibert, Y.},{Chabrier, G.},{Benz, W.}}{Baraffe et~al.}{2006}]{Baraffe06}
          {Baraffe, I.},{ Alibert,Y.},{ Chabrier, G.},{Benz, W. }, 2006, \aap, 450, 1221-1229
          %Cited
          
          
            %Cited
  \bibitem[\protect\citeauthoryear{{Benz, W.},{ I da, S.},{ Alibert, Y.},{ Lin, D.},{ Mordasini, C.}}{{Benz et~al.}}{2014}]{Benz14}
  {{Benz, W.},{ Ida, S.},{ Alibert, Y.},{ Lin, D.},{ Mordasini, C.}}, 2014, Protostars and planets VI  %Cited
        
  
\bibitem[\protect\citeauthoryear{{Bodenheimer} \& {Lissauer}}{{Bodenheimer} \&
  {Lissauer}}{2014}]{BodLiss14}
{Bodenheimer, P.} \& {Lissauer, J.~J.}, 2014, \apj, 791, 103

\bibitem[\protect\citeauthoryear{{Bodenheimer} \& {Pollack}}{{Bodenheimer} \&
  {Pollack}}{1986}]{BP86}
{Bodenheimer, P.} \& {Pollack}, J.~B., 1986, \icarus, 67, 391

 %Cited
 \bibitem[\protect\citeauthoryear{{Brouwers M. G. }, {A. Vazan},{C. W. Ormel}}{{Brouwers}	et~al.}{2017}]{Brouwers2017}
 {Brouwers, M. G.},{ A. Vazan,}{ C. W. Ormel}, 2017, Astronomy and Astrophysics, Forthcoming article
 %Cited 
 
  %Cited
 \bibitem[\protect\citeauthoryear{{Chyba C. F.}, {Thomas P. J.},{Zahnle K. J.}}{{Chyba}	et~al.}{1993}]{Chyba93}
 {Chyba, C. F.,}{ Thomas, P. J.,}{ Zahnle, K. J.}, 1993, Nature 361, 40-44
 %Cited 
 
 %Cited
 \bibitem[\protect\citeauthoryear{{Fortney, J. J.}, {Mordasini C.},{Nettelmann, N.},{Kempton, E.},{Greene, T.},{Zahnle, K.}}{{Fortney}	et~al.}{2013}]{Fortney13}
 {Fortney, J. J.,}{ Mordasini, C.,}{ Nettelmann, N.,}{ Kempton, E.,}{ Greene, T.,}{ Zahnle, K.,} 2013, \apj, 775, 1
 %Cited 

 %Cited
\bibitem[\protect\citeauthoryear{{Freedman, R. S. }, {Lustig-Yaeger, J. },{Fortney, J. J}}{{Freedman}	et~al.}{2014}]{Freedman14}
{Freedman, R. S.},{ Lustig-Yaeger, J.,}{ Fortney, J. J,}  et al. 2014, \apjs, 214, 25
%Cited 

\bibitem[\protect\citeauthoryear{{Hayashi}}{{Hayashi}}{1981}]{Hayashi81}
{Hayashi}, C., 1981, Progress of Theoretical Physics Supplement, 70, 35

\bibitem[\protect\citeauthoryear{{Helled} et~al.}{{Helled}
  et~al.}{2011}]{Helled11}
{Helled, R.},{ Anderson, J.~D.},{ Podolak, M.},{ Schubert, G.}, 2011, \apj, 726, 15

\bibitem[\protect\citeauthoryear{{Helled} \& {Bodenheimer}}{{Helled} \&
  {Bodenheimer}}{2014}]{HelledBod14}
{Helled}, R.,  \& {Bodenheimer}, P., 2014, \apj, 789, 69

  %Cited
  \bibitem[\protect\citeauthoryear{{Helled, R.},{Lunine, J.}}{{Helled \& Lunine}}{2014}]{Helled14}
  {{Helled, R.},{ Lunine, J.}}, 2014, Monthly Notices of the Royal Astronomical Society, 441(3)
  %Cited

  \bibitem[\protect\citeauthoryear{{Helled} et~al.}{{Helled}
  et~al.}{2014}]{HelledPPVI}
  {Helled, R.} et~al., 2014, Protostars and Planets VI, 643

 %Cited
  \bibitem[\protect\citeauthoryear{{Helled R.},{David Stevenson}}{{Helled \& Stevenson}}{2017}]{Ravit17}
  {Helled, R.},{ Stevenson David}, 2017, \apj, 840, 1
   %Cited
            
   %Cited
   \bibitem[\protect\citeauthoryear{{Helled R.},{M Podolak},{A. Kovetz}}{{Helled  et~al.}}{2006}]{Ravit06}
   {Helled, R.},{ Podolack, M.},{ Kovetz, A.}, 2006, Icarus, 185, 1
   %Cited
            	        
   %Cited
  \bibitem[\protect\citeauthoryear{{Hills, J. G.}, {Goda, M. P.}}{{Hills \& Goda}}{1993}]{Hills93}
  {Hills, J. G.}, {Goda, M. P.}, 1993, \aj, 105, 3
  %Cited

  \bibitem[\protect\citeauthoryear{{Hori} \& {Ikoma}}{{Hori} \& {Ikoma}}{2011}]{HI11}
  {Hori}, Y. \& {Ikoma}, M., 2011, \mnras, 416, 1419

  \bibitem[\protect\citeauthoryear{{Iaroslavitz} \& {Podolak}}{{Iaroslavitz} \&
  {Podolak}}{2007}]{Iaroslav07}
  {Iaroslavitz}, E. \& {Podolak}, M., 2007, \icarus, 187, 600

  \bibitem[\protect\citeauthoryear{{Ida} \& {Makino}}{{Ida} \&
  {Makino}}{1993}]{IdaMak93}
  {Ida}, S. \& {Makino}, J., 1993, \icarus, 106, 210

  \bibitem[\protect\citeauthoryear{{Ikoma} \& {Hori}}{{Ikoma} \&
  {Hori}}{2012}]{IkomaHori12}
  {Ikoma}, M. \& {Hori}, Y., 2012, \apj, 753, 66

  \bibitem[\protect\citeauthoryear{{Inaba} \& {Ikoma}}{{Inaba} \&
  {Ikoma}}{2003}]{Inaba03}
  {Inaba}, S. \& {Ikoma}, M., 2003, \aap, 410, 711

  \bibitem[\protect\citeauthoryear{{Lozovsky} et~al.}{{Lozovsky}
  et~al.}{2017}]{Lozovsky17}
  {Lozovsky, M.}, {Helled, R.}, {Rosenberg, E.~D.}, {Bodenheimer, P.}, 2017, \apj, 836, 227
  

  \bibitem[\protect\citeauthoryear{{Makinde O.D. },{Khan W.A. },{Khan Z.H. }}{{Makinde et~al.}}{2013}]{Makinde13}
  {Makinde, O.D.,}{ Khan, W.A., }{Khan, Z.H.,} 2013, International Journal of Heat and Mass Transfer 62, 526-533
  
  \bibitem[\protect\citeauthoryear{{Mehta, P.M.}{Minisci, E.M.}{Vasile, M.}}{{Mehta} et~al.}{2017}]{Mehta17}
  {Mehta, P. M.,}{ Minisci, E. M.,}{ Vasile, M.,} 2017, 4th IAA PDC
  
    \bibitem[\protect\citeauthoryear{{Melosh, H. J.,}{Goldin, T. J.}}{{Melosh \& Goldin}}{2008}]{Melosh08}
 {Melosh, H. J.,}{ Goldin, T. J.,} 2008, in Lunar and Planetary Science Conference

  \bibitem[\protect\citeauthoryear{{Mordasini}}{{Mordasini}}{2014}]{Mordasini14}
  {Mordasini, C.} 2014, \aap, 572, A118

  \bibitem[\protect\citeauthoryear{{Mordasini}, {Alibert}, \& {Benz}}{{Mordasini}
  et~al.}{2006}]{Mordasini06}
  {Mordasini, C.,} {Alibert}, Y. \& {Benz}, W., 2006, in Tenth Anniversary of 51 Peg-b: Status of and prospects for hot Jupiter studies, ed. L.~{Arnold}, F.~{Bouchy} \& C.~{Moutou}, 84

  %Cited
  \bibitem[\protect\citeauthoryear{{Mordasini, C.},{Mollière, P.},{Dittkrist},{K.-M., Jin},{Y. Alibert}}{{Mordasini et~al.}}{2015}]{Mordasini2015}
  {{Mordasini, C.,} {Mollière, P.,} {Dittkrist,} {Jin, K., M.,}{Alibert, Y,}} 2015, International Journal of Astrobiology, 14
  %Cited 
  
  %Cited
  \bibitem[\protect\citeauthoryear{{Mordasini, C.},{Alibert, Y.},{Benz, W.}}{{Mordasini et~al.}}{2009}]{Mordasini2009}
   {{Mordasini, C.,}{ Alibert, Y.,}{ Benz, W.,}} 2009, \aap, 501, 3, July III
  %Cited 
  
   %Cited
    \bibitem[\protect\citeauthoryear{{Mordasini, C.},{Boekel, R.},{Mollière, P.},{Henning, Th.},{Benneke, B.}}{{Mordasini et~al.}}{2016}]{Mordasini16}
    {{Mordasini, C.,}{ Boekel, R.,}{ Mollière, P.,}{ Henning, Th.,}{ Benneke, B.,}} 2016, \apj, 832, 1
    %Cited 
  
    \bibitem[\protect\citeauthoryear{{Movshovitz} et~al.}{{Movshovitz}
    et~al.}{2010}]{Movshovitz10}
   {Movshovitz, N.,}{ Bodenheimer, P.,}{ Podolak, M.,}{ Lissauer, J.~J.,} 2010,
    \icarus, 209, 616

   \bibitem[\protect\citeauthoryear{{Movshovitz} \& {Podolak}}{{Movshovitz} \&
   {Podolak}}{2008}]{Movshovitz08}
   {Movshovitz, N.} \& {Podolak, M.,} 2008, \icarus, 194, 368
  
   \bibitem[\protect\citeauthoryear{{Nijemeisland} \& {Dixon}}{{Nijemeisland} \&
   {Dixon}}{2004}]{Nijemeisland04}
   Nijemeisland, M. \& Dixon, A. G., 2004, AIChE J., 50, 906-921.
  
   %Cited
   \bibitem[\protect\citeauthoryear{{ Paul J.Register},{Donovan L.Mathias},{Lorien F.Wheeler}}{{Register et~al.}}{2017}]{Register2017}
   {{Register, J. P.,}{ Mathias, D. L.,}{Wheeler, L. F.}} 2017 \icarus, 284, 157–166. 
  %Cited
  
   %Cited
   \bibitem[\protect\citeauthoryear{{Fausto Perri},{A.G.W. Cameron}}{{Perri} \& {Cameron}}{1974}]{PerriCameron1974}
   {Perri, F.},{ Cameron, A. G. W.,} 1974, \icarus, 22, 416
 %Cited 
  
   %Cited
    \bibitem[\protect\citeauthoryear{{Arazi Pinhas}, {Nikku Madhusudhan},{Cathie Clarke}}{{Pinhas}	et~al.}{2016}]{Pinhas2016}
    {Pinhas, A.,}{ Madhusudhan, N.},{ Clarke, C.,} 2016, \mnras, 463, 4, 21, 4516–4532
   %Cited  +
   


     %Cited
     \bibitem[\protect\citeauthoryear{{Pletcher H. Richard }, {Tannehill C. John  }, {Anderson Dale }}{{Pletcher et~al.}}{2012}]{Pletcher12}
     {{Pletcher, R. H.,}{ Tannehill, J. C.,}{ Anderson, D.,}} 2012, Computational fluid mechanics and heat transfer, third edition
    %Cited
  
    \bibitem[\protect\citeauthoryear{{Podolak}}{{Podolak}}{2003}]{Podolak03}
    {Podolak}, M. 2003, \icarus, 165, 428

    \bibitem[\protect\citeauthoryear{{Podolak}, {Pollack}, \& {Reynolds}}{{Podolak}
    et~al.}{1988}]{Podolak88}
    {Podolak, M.},{ Pollack, J.~B.},{ Reynolds, R.~T.}, 1988, \icarus, 73, 163

    \bibitem[\protect\citeauthoryear{{Pollack} et~al.}{{Pollack}
    et~al.}{1996}]{P96}{Pollack, J.~B.},{ Hubickyj, O.,}{ Bodenheimer, P.},{ Lissauer, J.~J.},{ Podolak, M.,}{ Greenzweig, Y.,} 1996, \icarus, 124, 62

    \bibitem[\protect\citeauthoryear{{Pollack} et~al.}{{Pollack}
	et~al.}{1986}]{P86}
    {Pollack, J.~B.},{ Podolak, M.,}{ Bodenheimer, P.,}{ Christofferson, B.} 1986, \icarus, 67, 3
  
  
    \bibitem[\protect\citeauthoryear{{Revelle, D. O.}}{{Revelle}}{2005}]{Revelle05}
    {Revelle, D. O.,} 2005, Earth, Moon, and Planets 95: 441-476
  
    \bibitem[\protect\citeauthoryear{{Revelle, D. O.}}{{Revelle}}{2007}]{Revelle07}
    {Revelle, D. O.}, 2007, Near Earth objects, our celestial neighbors: Opportunity and risk, 95-106
  
    %Cited
    \bibitem[\protect\citeauthoryear{{ Saumon, D.},{Chabrier, G.},{van Horn, H. M.}}{{Saumon et~al.}}{2015}]{Saumon1995}
    {{Saumon, D.,}{ Chabrier, G.,}{ van Horn, H. M.,}} 1995, \apjs, 99, 713
    %Cited 

  
    %Cited
     \bibitem[\protect\citeauthoryear{{ Svetsov , V. V.},{I. V. Nemtchinov},{ A. V. Teterv}}{{Svetsov et~al.}}{1995}]{Svetsov1995}
     {{Svetsov , V. V.,}{Nemtchinov, I. V.,}{Teterv, A. V.,}} 1995, \icarus, 116, 131–153.
     %Cited  
   
      %Cited
   \bibitem[\protect\citeauthoryear{{Allona Vazan}, {Ravit Helled }, {Tristan Guillot }}{{Vazan et~al.}}{2018}]{Vazan2018}
   {{Vazan, A.},{Helled, R. }, {Guillot, T.}}, 2018, \aap, 610, L14 
   %Cited
   
   %Cited
   \bibitem[\protect\citeauthoryear{{Allona Vazan},{Ravit Helled},{Morris Podolak},{ Attay Kovetz}}{{Vazan et~al.}}{2016}]{Vazan2016}
   {Vazan, A.}, {Helled, R.}, {Podolak, M.},{Kovetz, A.}, 2016, \apj, 829, 2
   %Cited
   
     \bibitem[\protect\citeauthoryear{{Venturini}, {Alibert}, \& {Benz}}{{Venturini}
     et~al.}{2016}]{Venturini16}
     {Venturini, J.,}{ Alibert, Y.,}{ Benz, W.,} 2016, \aap, 596, A90

  
     \bibitem[\protect\citeauthoryear{{Venturini} et~al.}{{Venturini}
     et~al.}{2015}]{Venturini15}
     {Venturini, J.}, {Alibert, Y.},{ Benz, W.},{Ikoma, M.}, 2015, \aap, 576, A114
  
     \bibitem[\protect\citeauthoryear{{Wahl S. M.}{Hubbard W.B.}{Militzer B.}{Guillot T.}{Miguel Y.}{Movshovitz N.}{Kaspi Y.}{Helled R.}{Reese D.}{Galanti E.}{Levin S.}{Connerney J.E.}{Bolton S.J.}}{Wahl et~al.}{2017}]{Wahl2017}
     {Wahl et~al.,} 2017, Geophys. Res. Lett. 44, 4649–4659
     %Cited 

     %Cited
     \bibitem[\protect\citeauthoryear{{Weibull}}{1951}]{Weibull1951}
     {Weibull}, 1951, Appl. Mech., 10, 140-147
     %Cited 
  
     %Cited
     \bibitem[\protect\citeauthoryear{{ Wuchterl, G.}}{{ Wuchterl, G.}}{1993}]{Wuchterl1993}
     {Wuchterl, G.}, 1993, \icarus, 106, 323
     %Cited 
  
     %Cited
     \bibitem[\protect\citeauthoryear{{Zahnle}}{{Zahnle}}{1992}]{Zahnle92}
     {Zahnle, K., J.}, 1992, JGR, 97, 10243
  %Cited 
    

\end{thebibliography}
\end{document}